# Changes in R Leonis over Two Centuries: Period evolution and dust shell development

Mike Goldsmith[1]
*Independent researcher*

24th December 2025

## ABSTRACT

The AAVSO-based historical light curve of the oxygen-rich Mira variable R Leonis is used to determine and analyse the properties of the star's maxima and minima. The pulsation period is found to have shortened by about 3 days over the past two centuries. Superimposed on the mean period are clear modulations on timescales of approximately 35 and 98 years.

The depths of extrema show non-random behaviour: minima in particular display strong similarity in depth over extended intervals, with a mean depth difference of 0.26 magnitude for adjacent minima compared to a 0.38 magnitude difference for random pairs. Depth coherence persists for up to about 43 years. Examination of extrema depths over the full observational history of the star reveals gradual long-term changes, consistent with evolution of the circumstellar dust environment.

## 1. INTRODUCTION

R Leo is a Mira variable with a complex circumstellar environment of dust and gases. Like all Miras, it is subject to variations in the brightnesses and timings of its maxima and minima.

These variations arise from a complex interplay of photospheric and circumstellar processes. Pulsations of the photosphere, regulated by the κ-mechanism, generate shock waves that trigger the condensation of molecular species (notably TiO, VO, and $H_2O$) in the upper atmosphere, as well as the formation of dust particles composed of alumina, silicates, and iron oxides. The molecular opacity and dust together substantially augment the decline in light caused by the photospheric pulsations. As the photosphere subsequently reheats and contracts, much of this material is destroyed, while a fraction of the dust survives and is accelerated outwards by radiation pressure (see, for example, Ireland et al. 2006; Wittkowski et al. 2007). As a result, R Leo typically varies in brightness by about 4 magnitudes at V, with a substantial fraction of this amplitude arising from phase-dependent circumstellar opacity rather than photospheric changes alone.

---

[1]contact: mikegoldsmith639@gmail.com

In principle one could infer details of these processes by analyses of multi-wavelength light curve variations together with spectroscopic data, using the observed characteristics of individual cycles of maxima and minima to constrain model atmosphere parameters. However, while such analyses have proved very fruitful, detailed quantitative modelling of the atmospheric dynamics remains a challenge (largely because of the strongly nonlinear role of convection in energy transport).

Additionally, many of the known behaviours of Mira variables play out over many cycles: in particular, the circumstellar dust environments of the stars are known to develop over periods of many decades (see, for example, Whitelock 1999).

High-resolution observations in recent years, particularly with facilities such as the Atacama Large Millimeter/sub millimeter Array (ALMA), have revealed multiple, spatially resolved dust and molecular shells extending over several stellar radii (e.g. Hoai et al. 2023; Fonfría et al. 2019).

Fortunately, as only the fifth variable star to be discovered (by J A Koch in 1782), hundreds of thousands of photometric observations of R Leo are available, charting hundreds of maxima and minima. Of course, the quality and quantity of observations vary widely, and the vast majority are visual estimates. Nevertheless, with so much data, changes in many characteristics of the light curve can be determined, and statistical analyses of them can be used to distinguish stochastic variations from progressive ones and to determine parameter values and ranges.

In particular, these changing characteristics can be used to develop qualitative accounts of likely developments in the circumstellar environment of the star over the last two centuries, as well as of changes in pulsation pattern over that period.

In 1986, in just such a project, G R Hoeppe collated and analyzed all available visual observations of the star (Hoeppe 1987). With a period of about 312 days, R Leo had by then undergone 268 cycles, though the first to be observed well enough to establish maximum and minimum values was the $79^{th}$, in 1824.

Hoeppe noted that "although the distribution of the brightness [of minima] is roughly Gaussian, a nonparametric median test … was also applied to the data … the test statistic T exceeded in all cases the critical value of the chi-squared distribution at the 99.9% confidence level. … [for] the maximum brightnesses no variability could be detected at the

95% confidence level. This is probably due to the enormous scatter which prevents possible changes to become statistically significant."

Hoeppe did not suggest an explanation of the pattern of the minima, noting only that "there are long-term variations of the minimum brightness for which no satisfactory explanation is given by recent evolutionary models."

By 2025, R Leo had undergone 313 cycles since discovery, the majority of which have been observed by amateur or professional astronomers. The work described in this paper was undertaken to further study changes in the timing and magnitudes of the star's maxima and minima and if possible to account for them.

There are a number of challenges to such a project: (i) as a member of a Zodiacal constellation, R Leo is lost in the light of the Sun for several weeks each year, so numerous maxima and minima are unobservable; (ii) the large number of observations by so many observers with a range of equipment means that there is enormous scatter in the reported magnitudes (exacerbated by the fact that, as a red star whose brightness is being judged against much less red comparison stars, magnitude estimation can be very challenging); (iii) variations between different maxima or minima are only of the order of a few tenths of a magnitude, similar to the typical observer uncertainty.

## 2. EXTRACTION AND COLLATION OF EXTREMA VALUES

The primary source of data from the late 19[th] century onwards is the extensive database of the AAVSO[2] which, when extracted for this study, contained 108,315 observations of R Leo.

This monumental collection is a tribute to the hard and diligent work of hundreds of observers, and to the work of the AAVSO itself, which has collated them and made them readily and freely available to all.

From this dataset, all observations of U, B, R, and infrared magnitudes, together with "brighter than"/"dimmer than" values and duplicates were removed. Only visual estimates were used. See Appendix A for the background to this.

434 observations of R Leo made by Prof. Adalbert (Vojtěch) Šafařík between 1877 and 1892 were added to the dataset after being checked for consistency (see Appendix B).

To extract minima and maxima magnitudes from this AAVSO + Šafařík database, trials were made to fit the light curve in the vicinity of the lowest points with a range of curves (selected purely on geometrical considerations). It was found that a curve defined by the sech-squared function:

$$f(t) = A\ sech^2((t-t_0)/w) + C$$

fits the light curve well when constrained to a region about 100 days long, centred on the apparently dimmest point. Goodness of fit was optimized through a chi-squared test. The values of $A$ and $t_o$ of the best-fit curve were then adopted as the magnitude and Julian Date (JD) of the minimum.

The resulting curves were checked visually and unsuccessful fits (usually those where there were no data close to the minimum point) were rejected. 150 minima fits remained after this.

From these preliminary best-fit curves, residuals for each observation were collated for each observer. From these, standard deviation-based uncertainties were calculated for each observer and sech-squared curves then refitted, weighted with these uncertainties. In cases where an observer made fewer than 10 observations, the global median residual value was used as the uncertainty. (In some cases, AAVSO observations include reported error values, but it was found that using these did not produce convincing results).

---

[2] https://apps.aavso.org/v2/data/search/photometry/?target=r+leo&start_date=1800-01-01&end_date=2025-10-28&observer=&obs_campaign=&submit=Search

The quality of the fits was assessed by eye and the results classified as "good" (110 minima) or "rough" (40 minima).

The above process was applied both to magnitude-based light curves and to flux based ones, and many of the remaining analyses in this paper were tested against both types and against curves fitted both with and without the observer uncertainty weighting. Magnitude-based uncertainty-weighted results proved superior throughout, although improvements were modest.

The shapes of the maxima of R Leo are more dissimilar, flatter, and less symmetric than the minima. No suitable parametric curve was found that could reliably model these wide-ranging forms, so a non-parametric approach was adopted.

All data were first converted from magnitudes to fluxes to allow linear averaging and robust statistical treatment.

In order to check that each observation was consistent with the star's behaviour at that time, a local robustness filter was first applied: for every observation, all points within a 10-day window were examined, and a local median flux and median absolute deviation (MAD) were computed. Observations that were clear outliers with respect to this local distribution were excluded. This step removed spurious measurements while preserving the full temporal resolution of the remaining data.

The filtered flux measurements were then grouped into bins 6 days wide in order to consolidate the data on a short, astrophysically reasonable timescale. Within each bin, a second round of MAD-based filtering was applied to remove any remaining discordant points, and the surviving fluxes were averaged to provide a single representative flux value for each bin. This stage reduced observer scatter and sampling irregularities without imposing any assumed functional form on the light-curve shape, while seeking to preserve sufficient data points for robust averaging.

Finally, these smoothed, binned flux values were examined within 4-day extremum windows in order to obtain a value for the brightest sustained state of the star rather than a single noisy measurement. The brightest of the representative points within each extremum window was adopted as the maximum brightness estimate for that cycle, and converted back to a maximum magnitude. This window was chosen to be short enough that any real changes in the brightness of the star would be significantly less than the likely uncertainties of most observers.

These results were examined visually, poor fits rejected and the remainder classified as "good" or "rough", resulting in 113 good fits and 42 rough fits.

In addition to the AAVSO data discussed so far, there are two other sources of values for the historical magnitudes and dates of R Leo minima: Hoeppe 1987 and Waagen et al 2010. The latter includes values for the magnitudes of 87 maxima and 87 minima from 1905 (cycle 173) to 1987 (cycle 269). Quality assessments are included.

The AAVSO method for determining maxima and minima used by Waagen et al. is that defined by Leon Campbell (Campbell 1926). For each star, an average light curve was calculated by combining many observed cycles from the early 20th century. For later observations, this mean curve was superimposed on each individual cycle, and the times and magnitudes of the curve with the best visual alignment adopted as those of the maximum and minimum.

Hence, while the same source of data is used by Waagen et al as for the current study, the method of extraction of maxima and minima differs in that Waagen et al. assume that the shape of the light curve is the same for all cycles. In the present work, only the general mathematical form is fixed, and then only for the near-minimum part of the light curve. The coefficients of the fitted curve are obtained separately for each minimum. For maxima, no mean curve of any kind is assumed. Also, for minima, the current work applies the observer-specific uncertainty weighting explained above.

Hoeppe collected all published maxima and minima data, giving magnitude values from 1824 (cycle 29) to 1984 (cycle 265) for 122 maxima and 98 minima. The original papers from which he draws the data include individual observations only rarely and it is frequently not clear from those papers exactly how the values for maxima and minima were derived. Hoeppe notes a few cases where values are uncertain.

Results from the three sources of maxima and minima vales were collated and the best available values for each extremum selected. This selection took into account the reported quality of the values and, where all three sources provided a value, the agreement between them. AAVSO light curves were used also as a sanity check on the Hoeppe values – in a few cases the latter were rejected as being inconsistent with the AAVSO light curve.

The range of maxima magnitudes found is 4.5 to 6.9 (mean: 5.7, median: 5.8). For minima, the range is 8.9 to 10.9 (mean: 10.0, median: 10.0).

The resulting values for 181 maxima and 138 minima are given in Tables 1 and 2 in Appendix C and are those on which the analyses in the remainder of this paper are based. ; Machine-readable versions are available on request.

## 3. TEMPORAL ANALYSES

## 3.1 (O-C) analysis

A linear least-squares fit to the observed timings of maxima and minima was used to compute initial O–C values (figures 1 and 2).

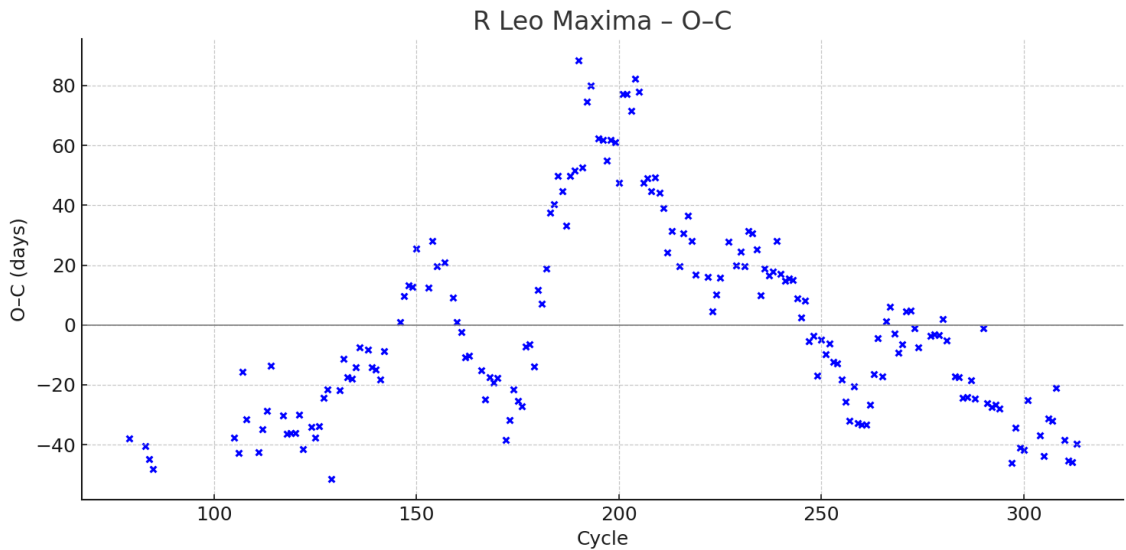

Figure 1 Differences between observed and calculated JDs of maxima.

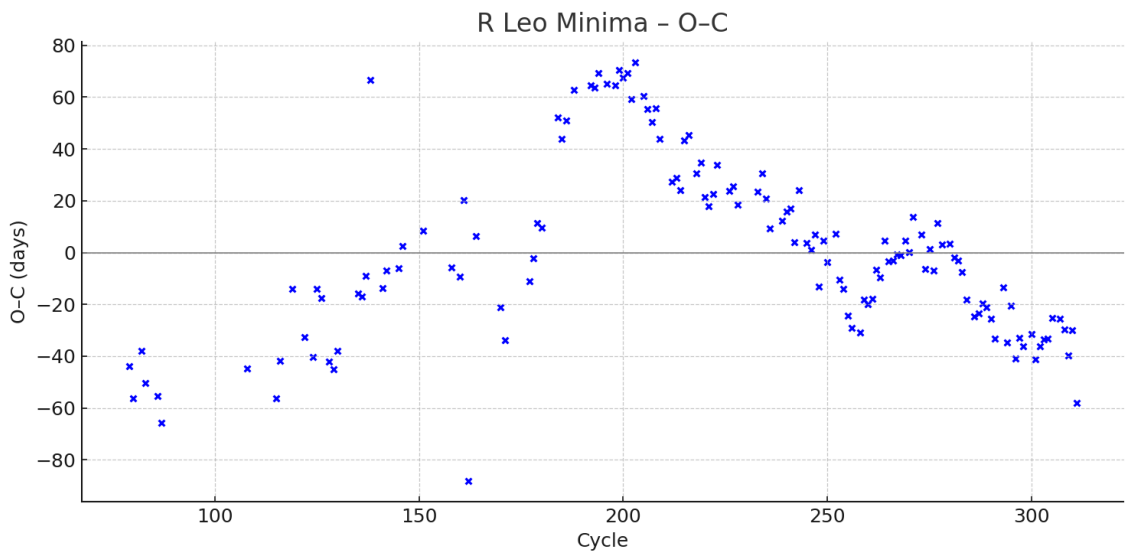

Figure 2 Differences between observed and calculated JDs of minima.

The general form of the plots is characteristic of a slow change in period: fitting and subtracting a quadratic curve (figures 3 and 4) confirmed a reduction in calculated mean period over the dataset.

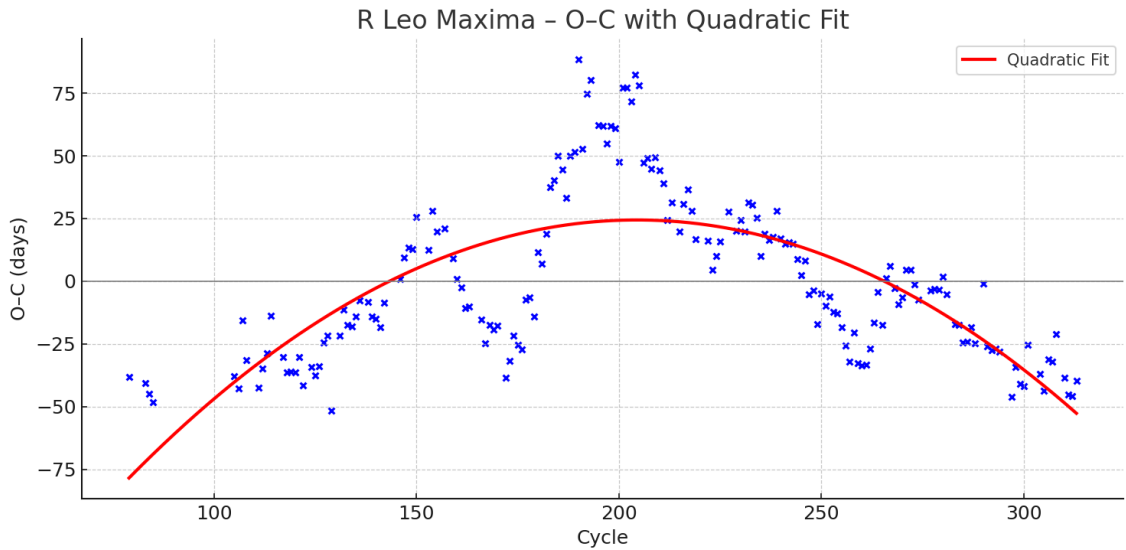

Figure 3: quadratic curve fitted to O-C values (maxima).

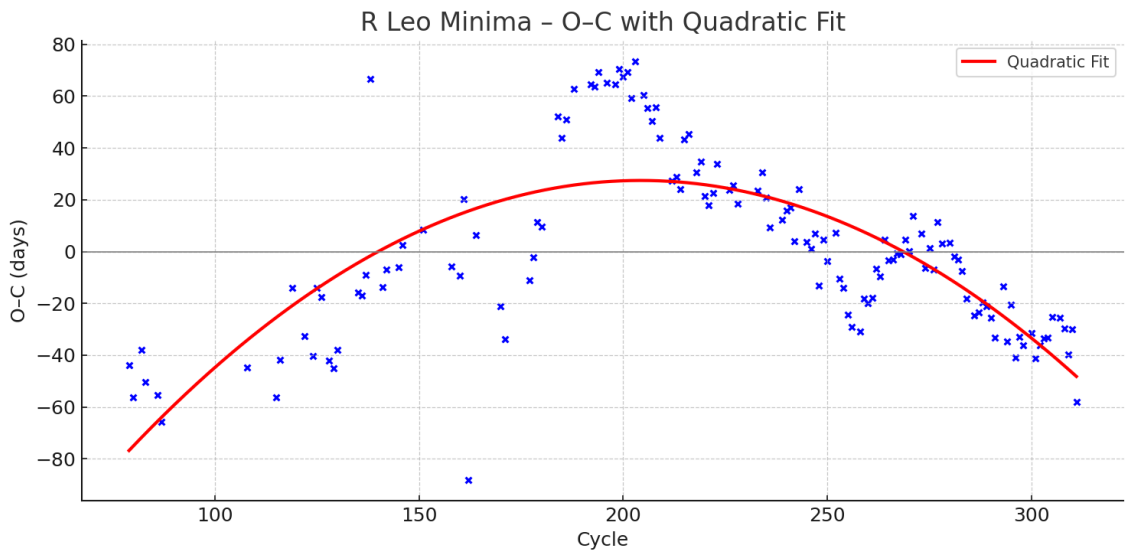

Figure 4: quadratic curve fitted to O-C values (minima).

Once the quadratic trend was removed, large deviations remained (figures 5 and 6).

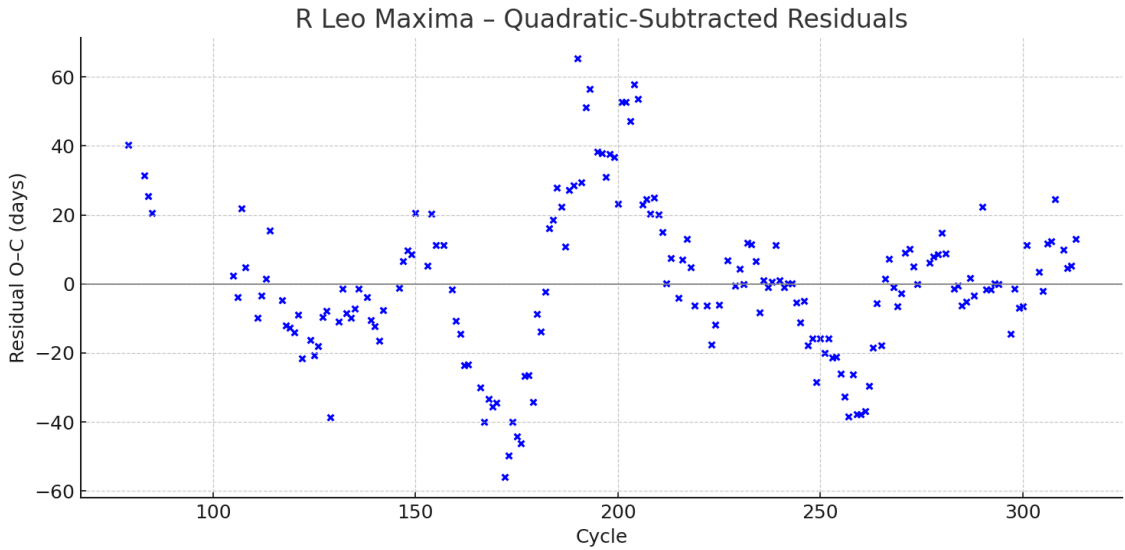

Figure 5: residual O-C values after subtraction of quadratic curve, maxima.

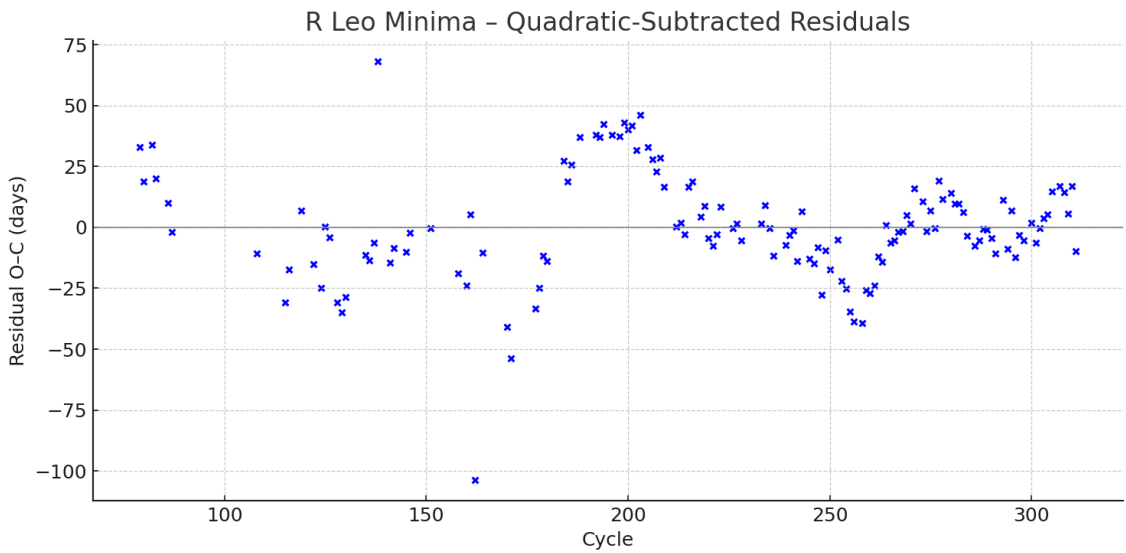

Figure 6: residual O-C values after subtraction of quadratic curve, minima.

Lomb–Scargle periodograms were applied to these residuals (figure 7).

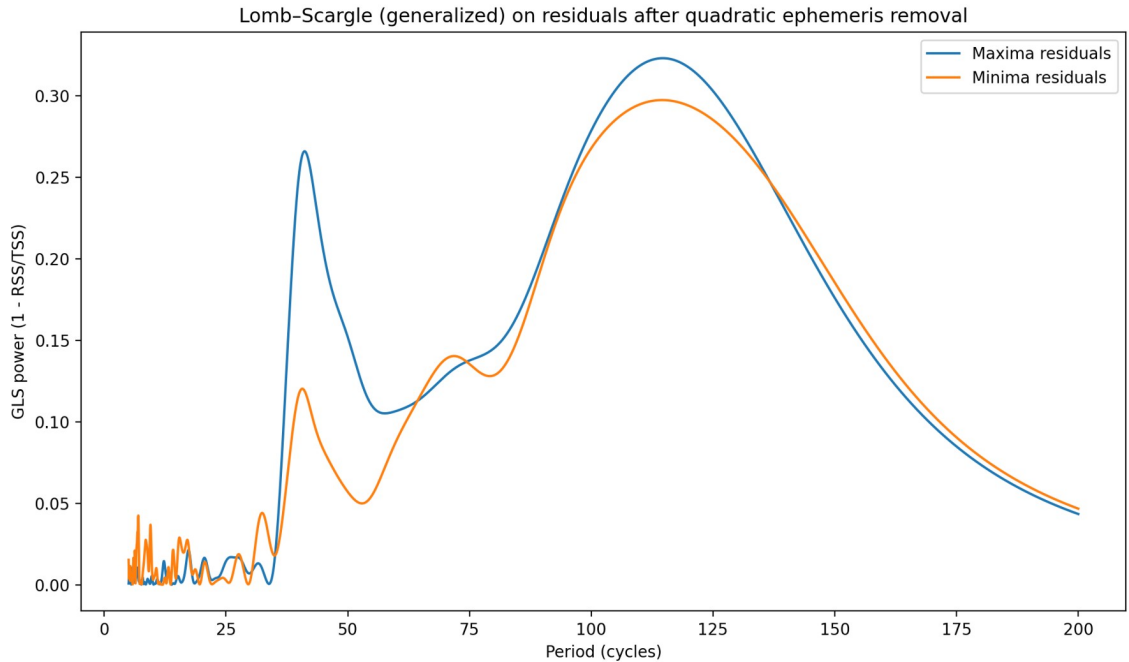

Figure 7: Lomb-Scargle periodograms of residuals.

The periodograms show that the timescales of phase modulation of the dominant ~312-day pulsation are about 41 and 115 cycles (about 35 and 98 years). It may be that the shorter variation is due to the disturbing effect of convective instabilities on the pulsation of the star. The longer timescale changes may be due to processes in the upper convective envelope, where dust and molecular species form. The weaker intermediate feature near 72 cycles (c. 62 years), seen most clearly in the minima, may arise from interaction between these processes.

However, with the longer timescale in particular, one should bear in mind that the observing history of R Leo is too brief to allow us to distinguish between periodic behaviour, stochastic shifts, or a stage of evolutionary development.

## 3.2 Ephemerides

The values of the ephemerides derived above are as follows. E is the Epoch (cycle number). All quoted periods and period changes are ephemeris-derived calculated (C) values. Deviations from these values will occur in individual observed cycles (O values) due to the intrinsic light-curve wandering mentioned above. For example, the observed separation (i.e., the O value) between minima 310 and 311 is 285 days, compared with a calculated local ephemeris period (C value) of ~311 days.

**Maxima**

**Centred Quadratic Ephemeris**

Cycles are measured relative to reference cycle $E_0 = 196$, i.e. $k = E - 196$.

$T(E) = (2424222.99 \pm 2.37) + (312.47 \pm 0.03) \cdot k - (0.00654 \pm 0.00045) \cdot k^2$

Period as a function of cycle: $P(E) = 312.47 - 0.01308 \cdot k$ days

Period change:
$(-4.19 \pm 0.29) \times 10^{-5}$ days per day
$-0.01308 \pm 0.00090$ days per cycle
$-1.53 \pm 0.11$ days per century

Total change across dataset (cycles 79–313; 201 years): $\Delta P\_total = -3.06 \pm 0.15$ days

Calculated recent mean period (at cycle 313): $\approx 310.94 \pm 0.11$ days

**Minima**

**Centred Quadratic Ephemeris**

Cycles are measured relative to reference cycle $E_0 = 195$, i.e. $k = E - 195$.

$T(E) = (2424085.55 \pm 2.86) + (312.48 \pm 0.03) \cdot k - (0.00664 \pm 0.00049) \cdot k^2$

Period as a function of cycle: $P(E) = 312.48 - 0.01328 \cdot k$ days

Period change:
$(-4.25 \pm 0.31) \times 10^{-5}$ days per day
$-0.01328 \pm 0.00098$ days per cycle
$-1.55 \pm 0.11$ days per century

Total change across dataset (cycles 79–311; 198 years): $\Delta P\_total = -3.08 \pm 0.17$ days

Calculated recent mean period (at cycle 311): $\approx 310.93 \pm 0.12$ days

## 3.3 Cycle lag persistence

The O–C value for each minimum is, on average, similar to that of the next minimum, the next-but one minimum, and so on. To determine how many minima this timing "memory" extends over, the Pearson correlation coefficient was computed for all pairs of minima separated by a fixed lag. See figure 8.

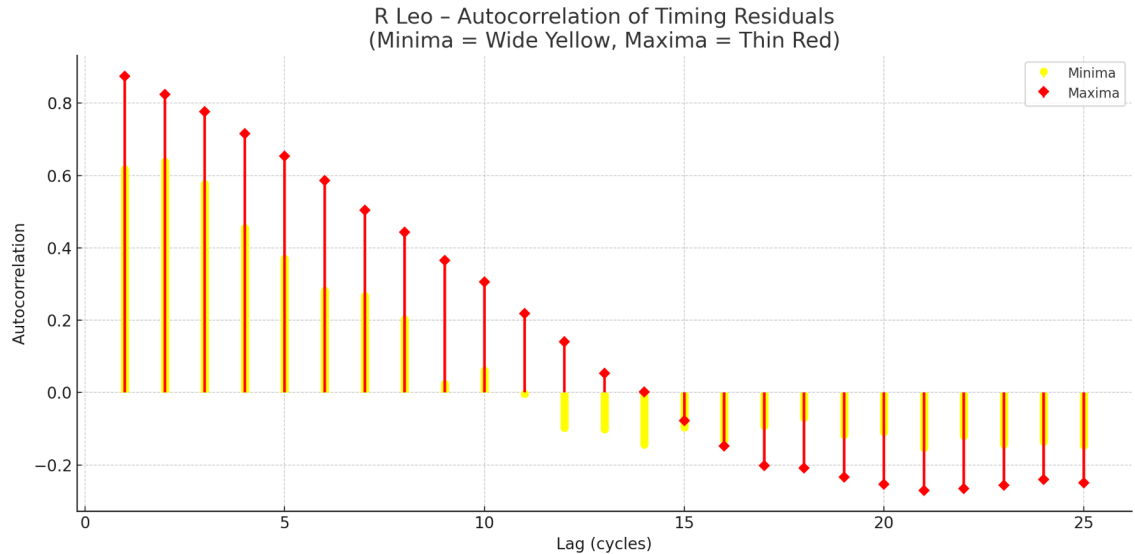

Figure 8: temporal memory.

The lagged correlation analysis shows that the timing residuals of both maxima and minima are non-random: early or late extrema tend to be followed by extrema which are also early or late. This behaviour persists for several cycles: for both minima and maxima the autocorrelation remains strongly positive out to lags of roughly 12-14 cycles (10-12 years). Beyond this, the correlation decays and becomes weakly negative.

In other words, for many cycles in a row, when the interval between minima or maxima is short, it tends to remain short, and when it is long, it remains long. The weak negative autocorrelation at larger lags is evidence that deviations are self-limiting, not cumulative; the star eventually returns to its long-term mean period.

The curves for maxima and minima are broadly similar, but the maxima correlation decays more smoothly, whereas the minima show slightly more structure and a somewhat sharper transition into the negative regime. This may be linked to the different ways the extrema are estimated: minima are timed using a common $\text{sech}^2$ profile and are often quite sharp, whereas maxima can be flatter and lack a single universal fitting function, so maxima timings may be more uncertain even though they are more numerous.

# 4. EXTREMA MAGNITUDES

## 4.1 Randomness tests

To test whether maxima and/or minima depths are randomly distributed, three independent statistical tests were applied, each sensitive to a different form of non-random structure:

**Wald–Wolfowitz runs test**
This non-parametric test examines the sequence of values above and below the median to determine whether their ordering is consistent with randomness. Clustering into long runs or excessively frequent alternation indicates a departure from random sequencing.

**Ljung–Box Q test**
This test evaluates whether a time-ordered series exhibits statistically significant autocorrelation. By testing multiple lags simultaneously, it probes whether values depend on earlier values rather than behaving as independent observations.

**Neighbour permutation test (mean squared neighbour differences)**
This test compares the observed average squared difference between successive data points with the distribution expected under random permutation. It is sensitive to both clustering (changes smaller than random) and over-dispersion (changes larger than random).

Results for maxima were as follows:

MAXIMA

The Ljung–Box Q test for autocorrelation yielded Q = 25.4 (approximate $\chi^2$ distribution with 10 degrees of freedom), corresponding to p ≈ 0.005, indicating a statistically significant departure from pure randomness. Because the Ljung–Box test is sensitive to correlations across multiple lags, this result suggests a possible degree of temporal "memory" in the sequence of maximum depths. However, when the test was repeated using only maxima classified as high quality, the p-value increased to approximately 0.27.

By contrast, neither of the other tests showed evidence for non-random structure.

The Wald–Wolfowitz runs test about the median gave a z-score of −0.52, with a two-sided p-value of approximately 0.60, consistent with random sequencing.

The permutation test on mean squared neighbour differences yielded an observed value of M_obs = 0.303, compared with a permutation mean of ⟨M⟩ = 0.308. The one-sided p-value for clustering was approximately 0.41, and the two-sided p-value was approximately 0.82, again indicating no departure from randomness.

MINIMA

The Ljung–Box Q test for autocorrelation yielded Q = 43.3 (approximate $\chi^2$ distribution with 10 degrees of freedom), corresponding to p ≈ 4 × 10$^{-6}$, indicating very strong evidence for autocorrelation in the sequence of minimum depths.

The Wald–Wolfowitz runs test about the median gave a z-score of −2.56, with a two-sided p-value of approximately 0.01, showing a statistically significant deviation from random sequencing.

The permutation test on mean squared neighbour differences yielded an observed value of M_obs = 0.168, compared with a permutation mean of ⟨M⟩ = 0.238. The one-sided p-value for clustering was p ≈ 2 × 10$^{-4}$, and the two-sided p-value was p ≈ 4 × 10$^{-4}$, indicating strong clustering of successive minima.

Taken together, the consistent rejection of the null hypothesis by all three tests provides strong evidence for non-random structure in the pattern of minima depths, involving both temporal correlation and clustering.

## 4.2 Periodicity tests

One possible explanation for these results is variability in photospheric brightness over a significantly longer timescale than the known c. 312 day period. To investigate this, Lomb–Scargle periodograms were constructed in cycle space, i.e. treating cycle number as the "time" variable and scanning for periods (4000 trial frequencies). The strongest peaks in the periodograms were tested using a permutation approach involving 1000 random shuffles of these magnitudes among the cycle numbers, for each of which the same periodograms were computed. The maximum powers from the shuffled datasets were compared with the maximum power of the actual data.

For the minima, the top five peaks in the periodogram are:
$P_1$ = 110.24 cycles, power ≈ 1.54
$P_2$ = 111.76 cycles, power ≈ 1.53
$P_3$ = 108.77 cycles, power ≈ 1.53
$P_4$ = 113.31 cycles, power ≈ 1.52
$P_5$ = 107.33 cycles, power ≈ 1.52

The probability of such results arising by chance is <0.1%. Hence the dominant structure is a broad, coherent bump around ≈110–115 cycles, rather than a single, sharply defined period.

For maxima the highest peaks in the Lomb–Scargle periodogram occur at:
$P_1$ = 2.82 cycles, power ≈ 1.12

$P_2$ = 2.81 cycles, power ≈ 1.11
$P_3$ = 2.83 cycles, power ≈ 1.11
$P_4$ = 2.84 cycles, power ≈ 1.10
$P_5$ = 2.80 cycles, power ≈ 1.10

However, the probability of these peaks arising by chance is 23%, so any significant periodicity or similar pattern cannot be supported by this analysis.

While the minima result is consistent with a secondary cycle of variability with a period of around a century, this is contradicted by the lack of such a pattern in the maxima.

In any case observations of R Leo have not yet been carried out long enough to trust such a conclusion: there has barely been time to observe two such cycles.

## 4.3 Tests on adjacent extrema

To study further the sense in which minima, and possibly maxima, have a "memory" of previous cycles, the differences in magnitude between adjacent extrema were compared with the differences between all pairs of extrema in the dataset. For each sequence (minima and maxima treated separately), the absolute magnitude difference was calculated for every adjacent pair, and the mean and median of these neighbour differences were determined. These were then compared with the corresponding mean and median magnitude difference taken over all possible pairs of extrema. In this way, any excess similarity between neighbours is measured directly as a deficit relative to the overall pairwise difference.

To evaluate whether such deficits might simply reflect chance fluctuations, a randomisation test was performed for each sequence: the list of magnitudes was shuffled many times, thereby destroying any temporal ordering, and for each shuffled version the same neighbour–difference statistics were recomputed. This generates a null distribution describing the range of neighbour differences expected if the extrema carried no memory of previous cycles.

For the minima, the mean absolute difference between neighbouring minima is 0.26 magnitude (100 adjacent pairs), compared with a global mean difference of 0.38 magnitude across all pairs of minima, giving a mean shift of −0.12 magnitude. The corresponding median values are 0.20 magnitude for neighbouring minima and 0.31 magnitude for all pairs, a median shift of −0.11 magnitude. So, adjacent minima are typically about a tenth of a magnitude closer to one another than two randomly chosen minima from the full record.

In the randomisation test, fewer than 1% of shuffled sequences produced neighbour differences as small as these.

See figures 9 and 10.

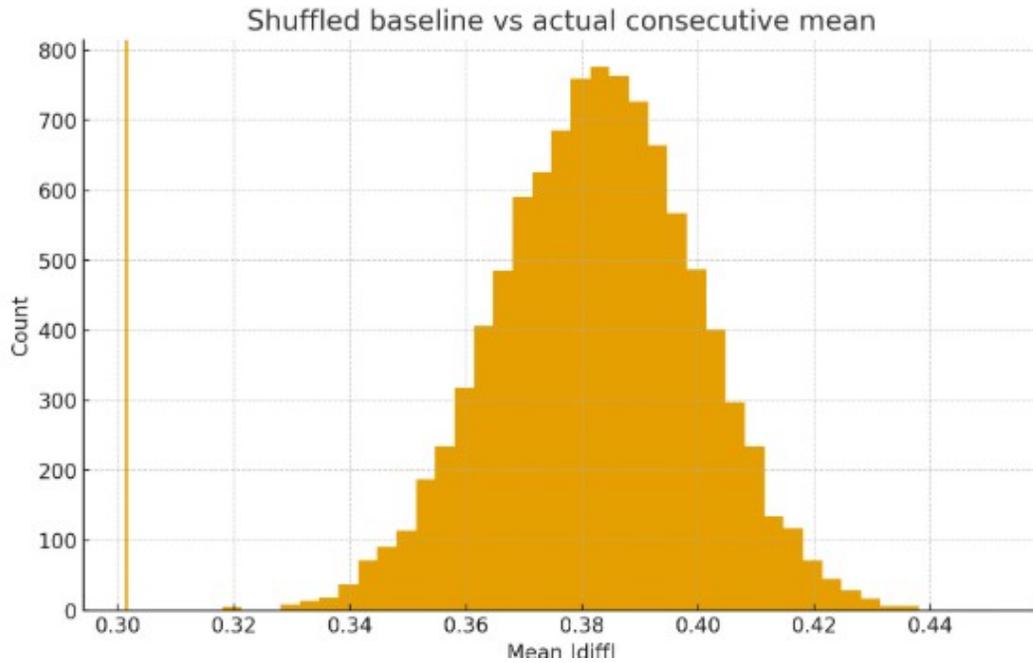

Figure 9. The vertical line is the mean difference of adjacent minima depths, the histogram shows the distribution of shuffled means. The probability of this occurring by chance is <0.0001.

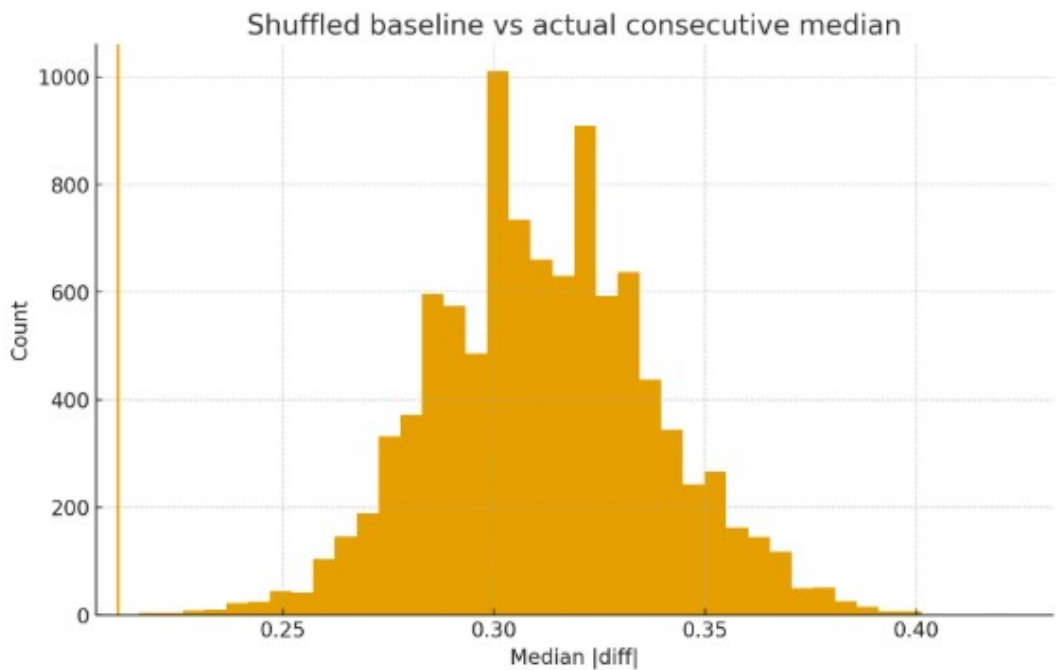

Figure 10. Median difference between depths of consecutive minima compared with histogram of median differences between random pairs.

For maxima, the effect is much weaker, despite there being many more pairs to test. The mean absolute difference between neighbouring maxima is 0.42 magnitude (156 adjacent pairs), compared with a global mean difference of 0.44 magnitude, a mean shift of only −0.02 magnitude. The median neighbour difference is 0.35 magnitude, compared with a global median of 0.39 magnitude, a median shift of about −0.03 magnitude. In this case, approximately 8–12% of the shuffled realisations yielded neighbour differences as small as the observed values. These small deficits are therefore entirely compatible with random variation, and any "memory" in the maxima is at best marginal, in contrast with the statistically significant effect seen in the minima.

(The reason that the mean difference between randomly selected maxima is nearly twice the mean difference for randomly selected minima is that maximum light in Miras is governed by highly variable shock amplitudes and a steep κ-opacity response, whereas minimum light is controlled by cooler atmospheric layers and slowly evolving molecular opacity, producing a narrower global distribution.)

It would be more satisfactory to have a clearer answer to the question of maximal memory. If there is such a memory, one might hypothesise that the same cause that leads to the similarity of adjacent minima would also lead to similarity of adjacent maxima. In this case, those pairs of cycles in which minima are especially close would presumably have especially close maxima also.

Not all cycles with known minima also have known maxima, nor is the converse true, so the domain of this test is rather limited: there are only 68 adjacent cycle pairs in which all four extrema are known.

Results are
Mean $\Delta(\min)$ between adjacent minima: 0.256.
Mean $\Delta(\max)$ between adjacent maxima: 0.406
Number of unusually-close-minimum pairs: 44
Number of pairs with both minima-close and maxima-close: 30
About 68% of cycles with unusually similar minima also have unusually similar maxima.
About 42% of random permutations produce at least 30 "close maxima" out of 44 draws.

Additionally a Pearson test for correlation between $\Delta\min$ and $\Delta\max$ gave r = 0.0748, with a probability of this result arising by chance of 54%, and a Spearman rank correlation test gave ρ = 0.0856, with a probability of this being chance of 49%

Hence, there is no evidence that adjacent cycles with minima closer than average also have maxima closer than average.

## 4.4 Decay of inter-minima similarity

Having established that minima tend to resemble their immediate neighbours with respect to depth, the next step is to determine how this similarity behaves as the separation between minima increases. To do this, every pair of minima with known depths was examined, regardless of how many cycles lay between them.

For each cycle separation *k* represented in the dataset, all pairs of minima that are exactly k cycles apart are identified, and the absolute magnitude differences for those pairs are computed. Averaging these differences for each *k* (using both the mean and the median) reveals how the typical separation between minima grows or diminishes as the gap between them widens.

To interpret these quantities, they were compared with the overall mean (or median) difference between all pairs of minima in the dataset, irrespective of their separation. This overall difference provides a baseline representing how far apart two minima would be "on average" if there were no relationship between them.

The shift function is defined as: $\Delta(k) = D(k) - D_{global}$ where $D(k)$ is the mean or median magnitude difference for minima separated by k cycles, and $D_{global}$ is the corresponding mean or median difference computed over all possible pairs of minima. $\Delta(k)$ therefore indicates whether minima k cycles apart are more similar than typical pairs ($\Delta(k) < 0$) or more different than typical pairs ($\Delta(k) > 0$). Plotting $\Delta(k)$ as a function of k shows how rapidly this "memory" fades as the cycle gap increases. Results are shown in figures 11 and 12.

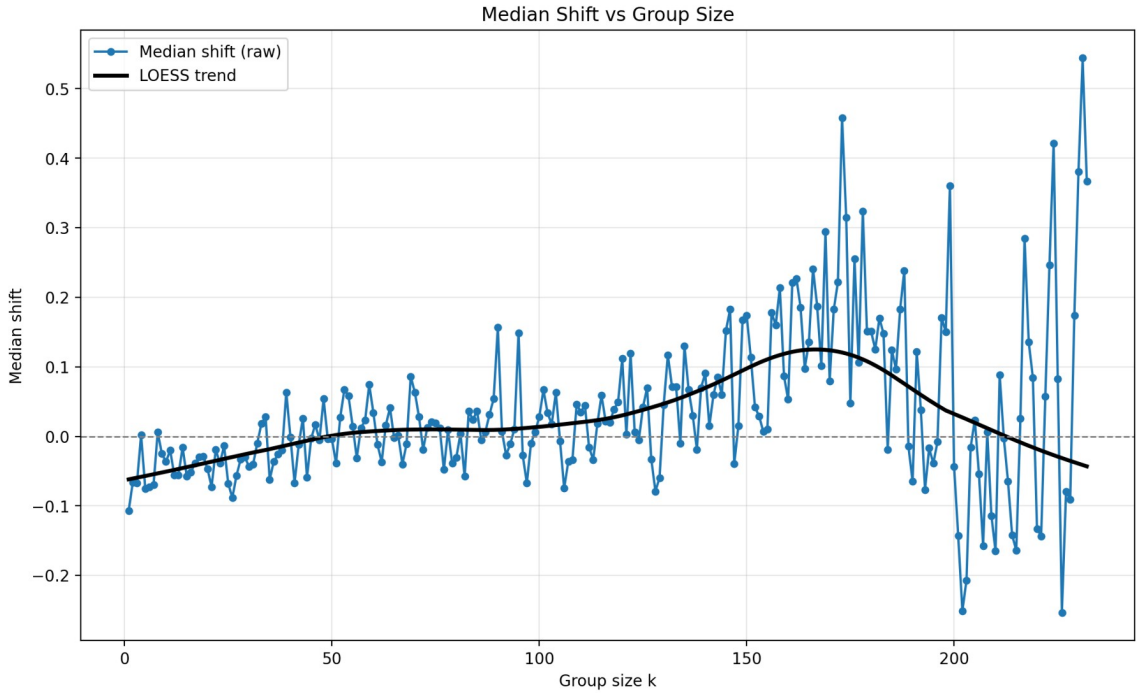

Figure 11. Median shift function vs. group size, with trend line.

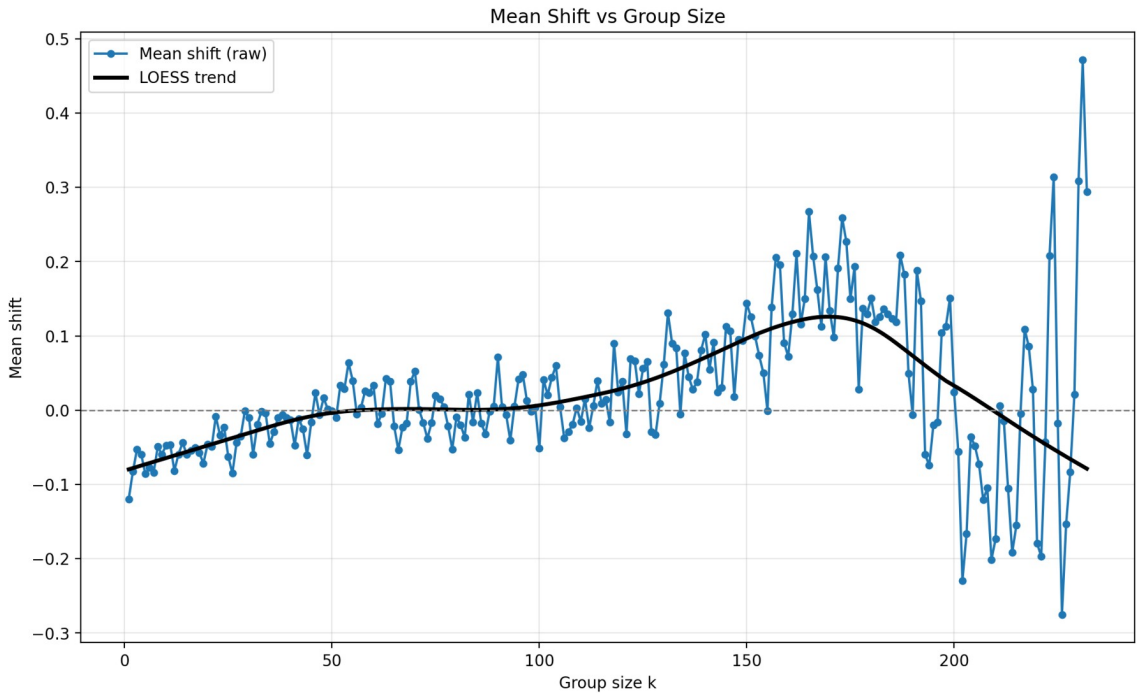

Figure 12. Mean shift function vs. group size, with trend line.

The plots of both the mean and median shifts show significant scatter, reflecting the intrinsic noise of visual magnitude estimates and the variation in quality and quantity of observations over the past two centuries. To illustrate the overall pattern a LOESS (locally weighted regression) smoothing is overlaid on each curve.

It is clear that the similarity in depth of nearby minima persists over several cycles, perhaps as many as 50.

Beyond this point, the smoothed curve remains close to zero for a substantial interval. Hence, over intermediate separations, the minima behave as if drawn from a stable underlying distribution with no persistent correlation between cycles.

At larger separations, a different behaviour emerges. After about 100 cycles, the smoothed curve rises above zero, reaching a broad maximum around cycle 170. Pairs separated by this many cycles are more different than typical pairs.

Beyond about cycle 180, the scatter increases dramatically, since the number of available pairs decreases sharply as $k$ approaches the total span of the dataset, and with so few points, noise dominates. The downturn of the LOESS curve in this region may therefore carry no physical significance.

One interpretation of these patterns is that they reflect timescales of change in the circumstellar environment, with dust/gas formation occupying up to 50 cycles (45 years) and destruction occurring over timescales greater than about 100 cycles (90 years).

However, since R Leo has only been observed for about 230 cycles (200 years), this must remain a speculative conclusion since so few such cycles could so far have occurred.

## 4.5 Mean magnitude over long periods

A disadvantage of the above analysis is that, while it shows the timescales over which R Leo's minima have darkened and brightened, it cannot show when these changes occurred – one pays for the statistical advantage of collating events over the history of observation of the star by sacrificing the sequence of those events.

Yet simply plotting all minima depths against time shows no pattern at all (figure 13).

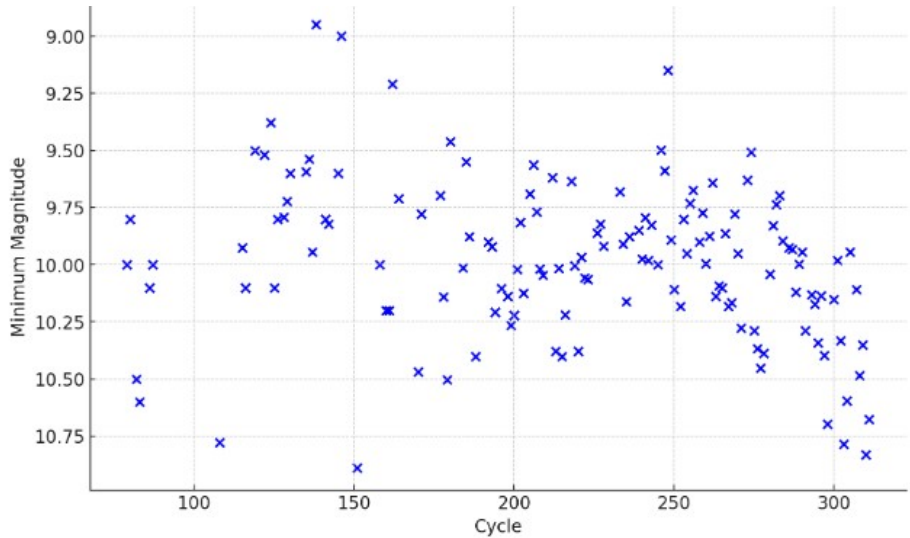

Figure 13. Sequence of known minima magnitudes.

Faced with this problem, Hoeppe used 5-cycle means but found no clear pattern, and revisiting this approach now, with the advantage of many more cycles of data, is still unenlightening (figure 14).

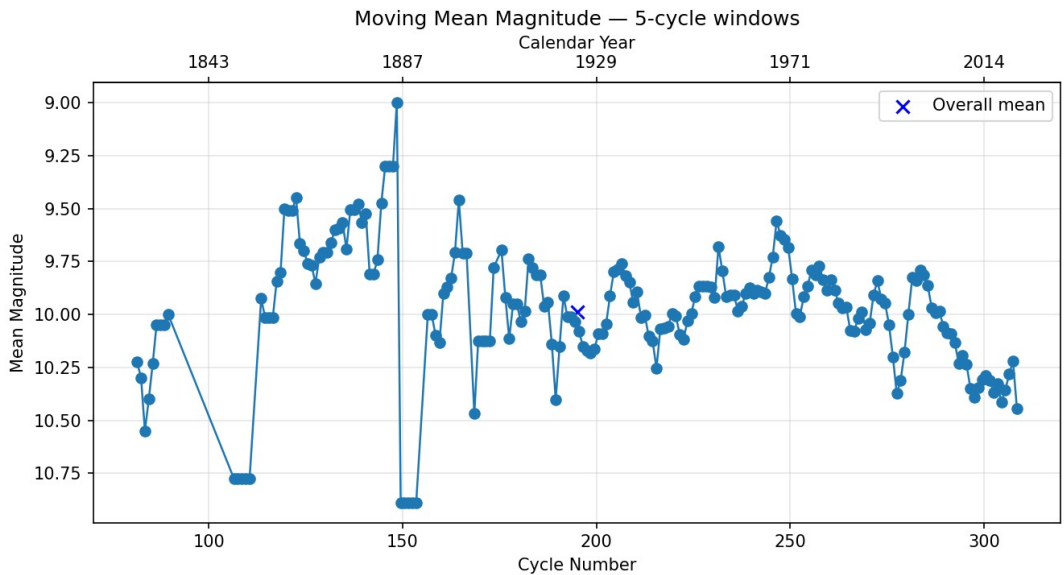

Figure 14. Five-cycle window mean minima magnitudes

Figure 14 was obtained by plotting the average of the magnitudes of the minima of the first five cycles, then the average of the second to the sixth, and so on.

However, the pairwise difference analysis in the previous section indicates that cycle-to-cycle coherence persists for up to about 50 cycles (≈43 years). When the size of the averaging window is increased to this scale, a much clearer structure emerges. Moving-mean curves computed with window lengths of this order naturally suppress short-term, cycle-to-cycle variations while retaining longer-timescale behaviour. Figure 15 shows the results of plotting the average magnitudes of groups of 50 cycles.

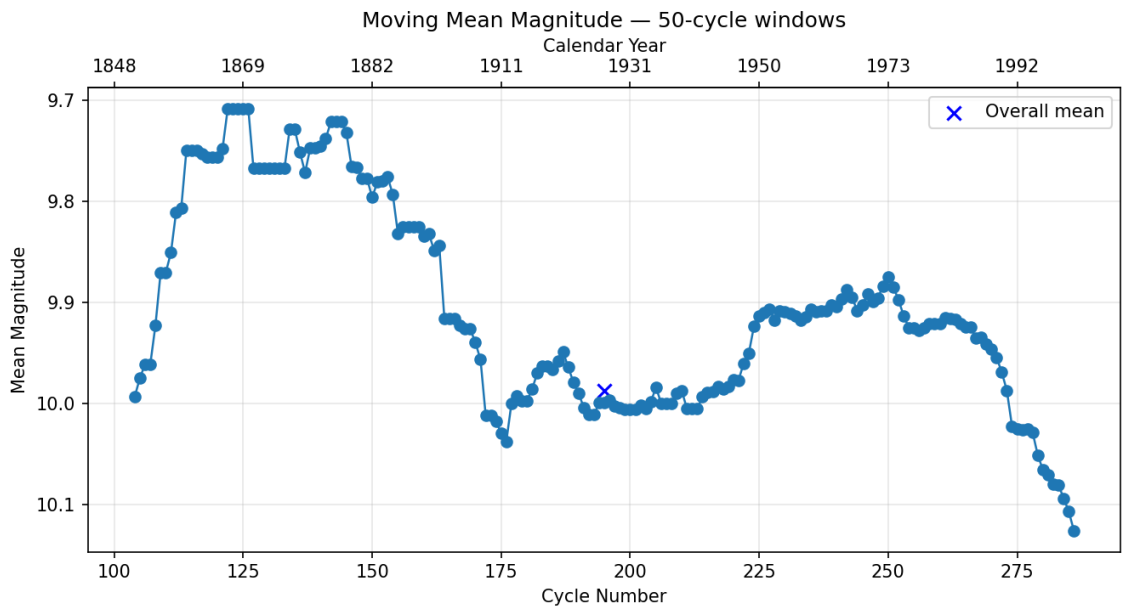

Figure 15 Fifty-cycle window mean minima magnitudes.

For the 50-cycle moving-mean curve, the median pointwise standard deviation across 2000 bootstrap resamples of the minima is ~0.05 magnitude. Hence, only features larger than this are likely to be real. These include a brightening of minima around the mid-19th century, followed by two extended declines: one from roughly the 1880s to about 1910, and another beginning in the mid-1970s. Between about 1860 and 1880 the minima remained comparatively bright and fairly stable, while from about 1910 to the mid-1970s they were again broadly stable but at a dimmer level. From about 1980 a steady reduction in minima brightness began.

The use of 50-cycle (~43 year) windows means that there are no mean values for the last and first c. 21 years of observations. A smaller window width salvages some of this region at the expense of a noisier profile (bootstrapped standard deviation is ~0.10 magnitude for 12-cycle windows). Figure 16 shows a continuing decline up until the end of the dataset, and reveals a possible sharp decline in the early 1840s, followed swiftly by an equally rapid recovery. Of course, such early observations are likely to be unreliable, especially since many early minima went unobserved. Nevertheless, the scale of these early changes, on the order of one magnitude, suggest that this change is real.

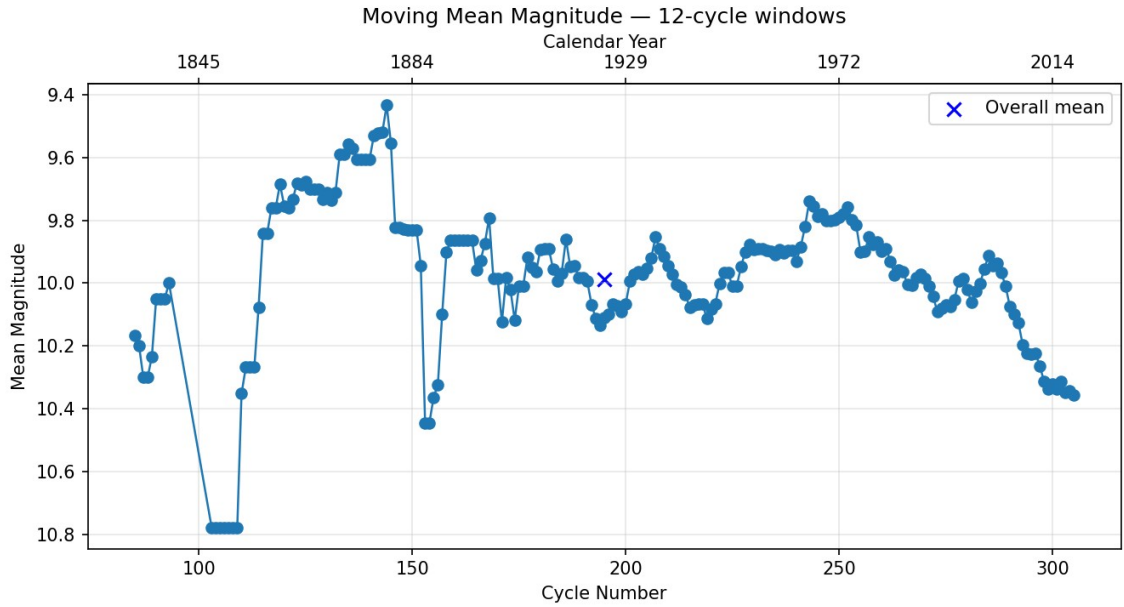

Figure 16 Twelve-cycle window mean minima magnitudes.

Overlaying the plots for results for windows sized from 5 to 168 cycles shows the persistence of the general form of the light curve for minima depths (figure 17).

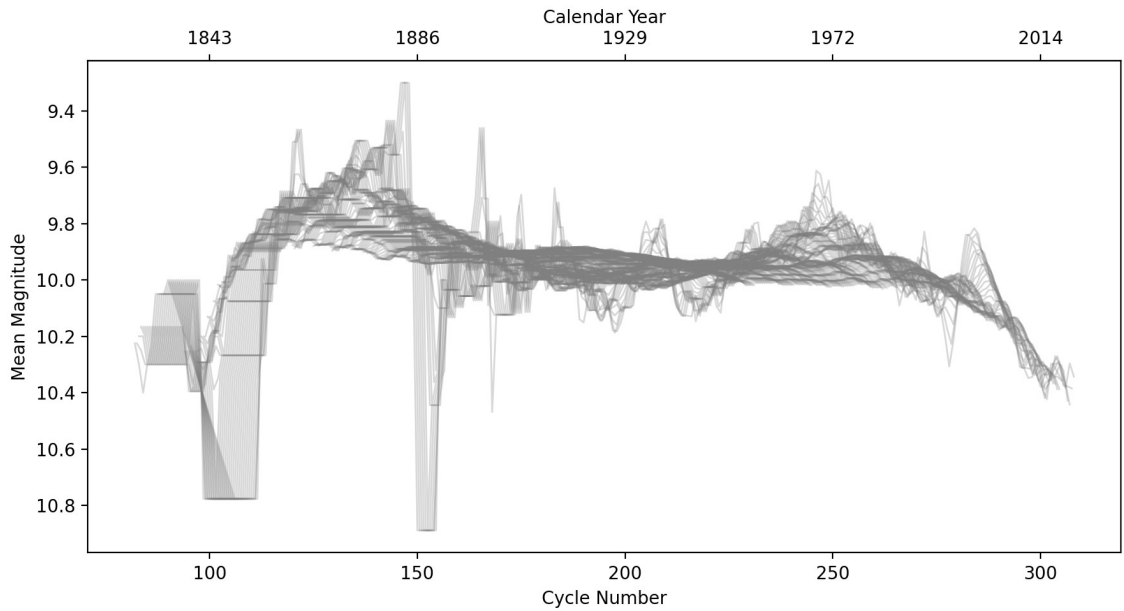

Figure 17 Five-cycle to 168-cycle window mean minima magnitudes.

For maxima, a broadly similar pattern emerges, but with a smaller magnitude variation (figure 18).

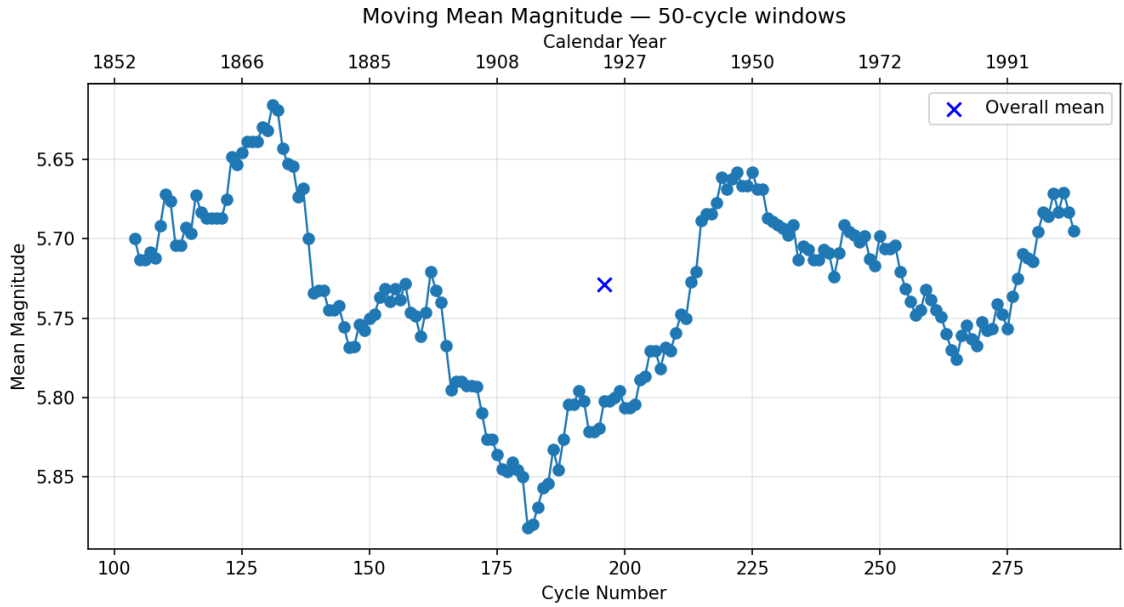

Figure 18 Fifty-cycle window mean maxima magnitudes. Bootstrapped standard deviation is ~0.06.

Reducing the window length results in a plot which is significantly contaminated by noise (figure 19). For 12-cycle windows, bootstrapped standard deviation is ~0.11.

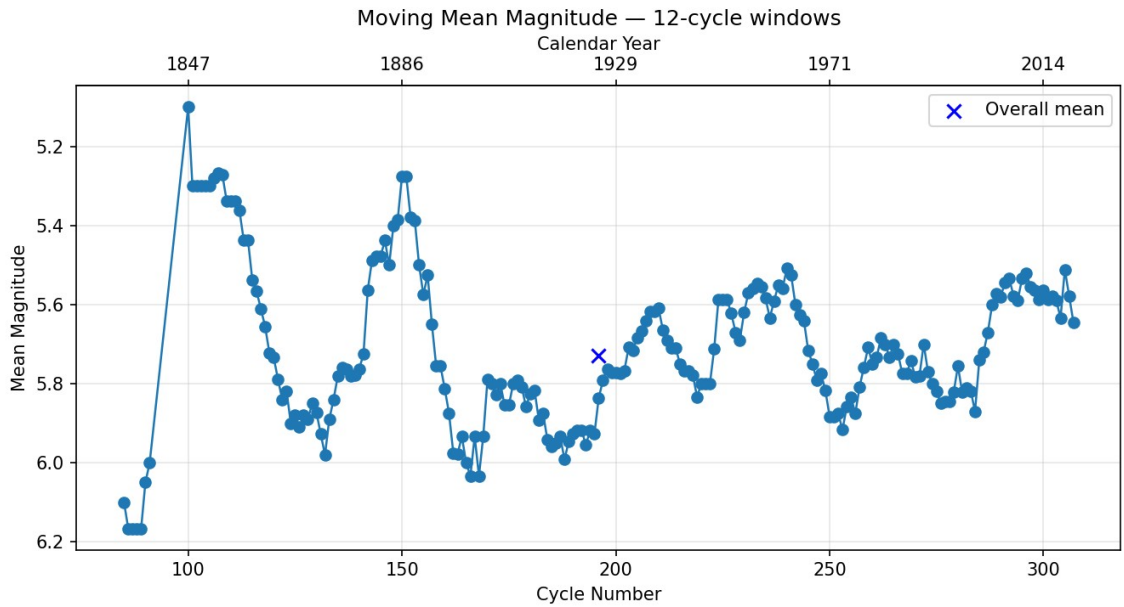

Figure 19 Twelve-cycle window mean maxima magnitudes.

Figure 20 compares 50 cycle moving mean magnitudes for both types of extremum.

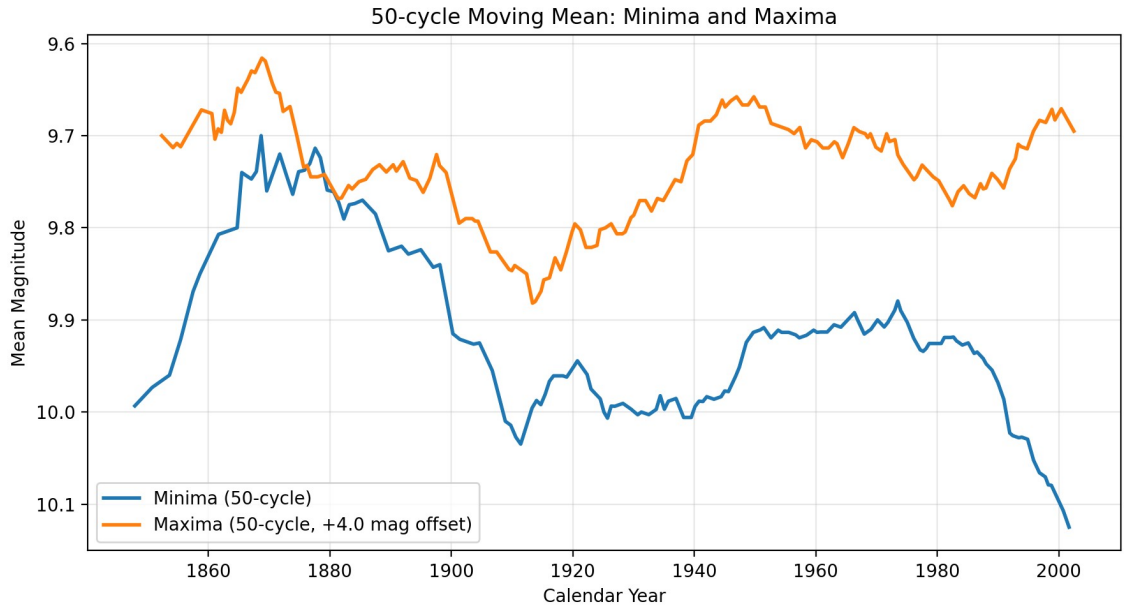

Figure 20 Comparison of moving means of maxima and minima.

While there is a general similarity in the shapes of the curves for maxima and minima, there is also considerable variation between them. This fact, and the general forms, of the curves make it plausible that changes in dust and/or molecular species rather than in stellar pulsation are the source of the alterations in extrema magnitudes.

One possibility is that R Leo occasionally produces abundant dust in its upper atmosphere over very short periods, much like an R Coronae Borealis star. This dust might then require multiple maxima to fully clear. However, if this were the case one would expect the minimum that accompanies each abrupt dust-forming event to be exceptionally deep. In fact, a histogram of all minima depths (figure 21) shows that no such minima have ever been recorded.

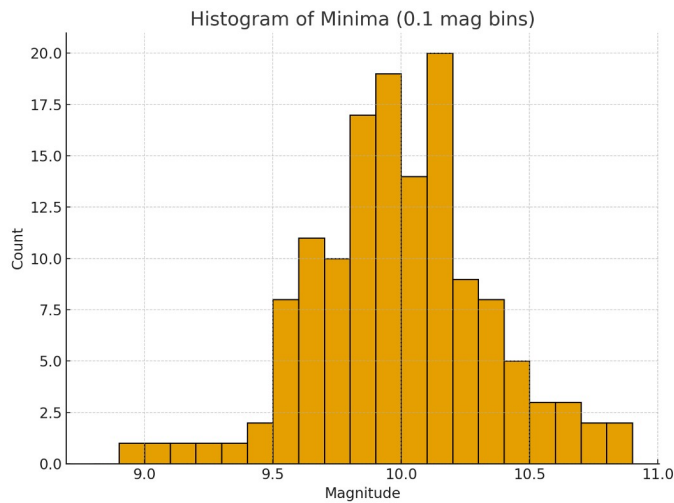

.
Figure 21 Distribution of depths of recorded minima.

## 5. RELATIONSHIPS BETWEEN DEPTHS AND TIMINGS OF EXTREMA

The existence of multi-cycle "memories" of both the periods and the depths of R Leo's minima prompts the question as to whether these are linked. One way to study this is to compare minimal depth with O-C value to establish whether, for example, more delayed minima tend to be darker (figure 22).

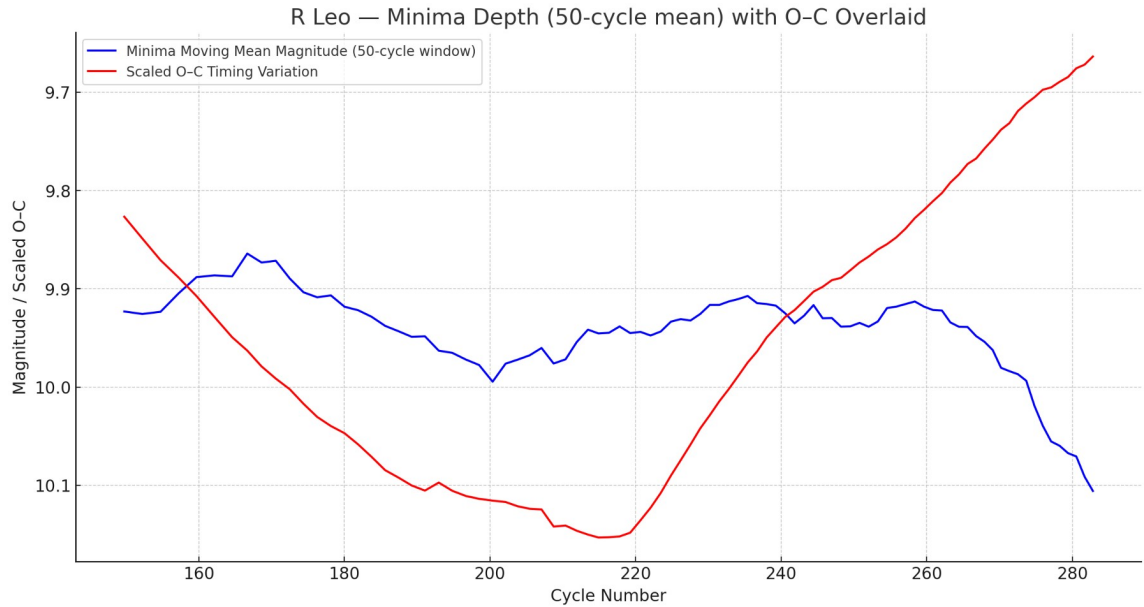

Figure 22 Comparison of change of minima depth with O-C value.

An alternative approach is to plot minima brightness against phase offset (i.e. O-C expressed as a fraction of period). Figure 23 shows this.

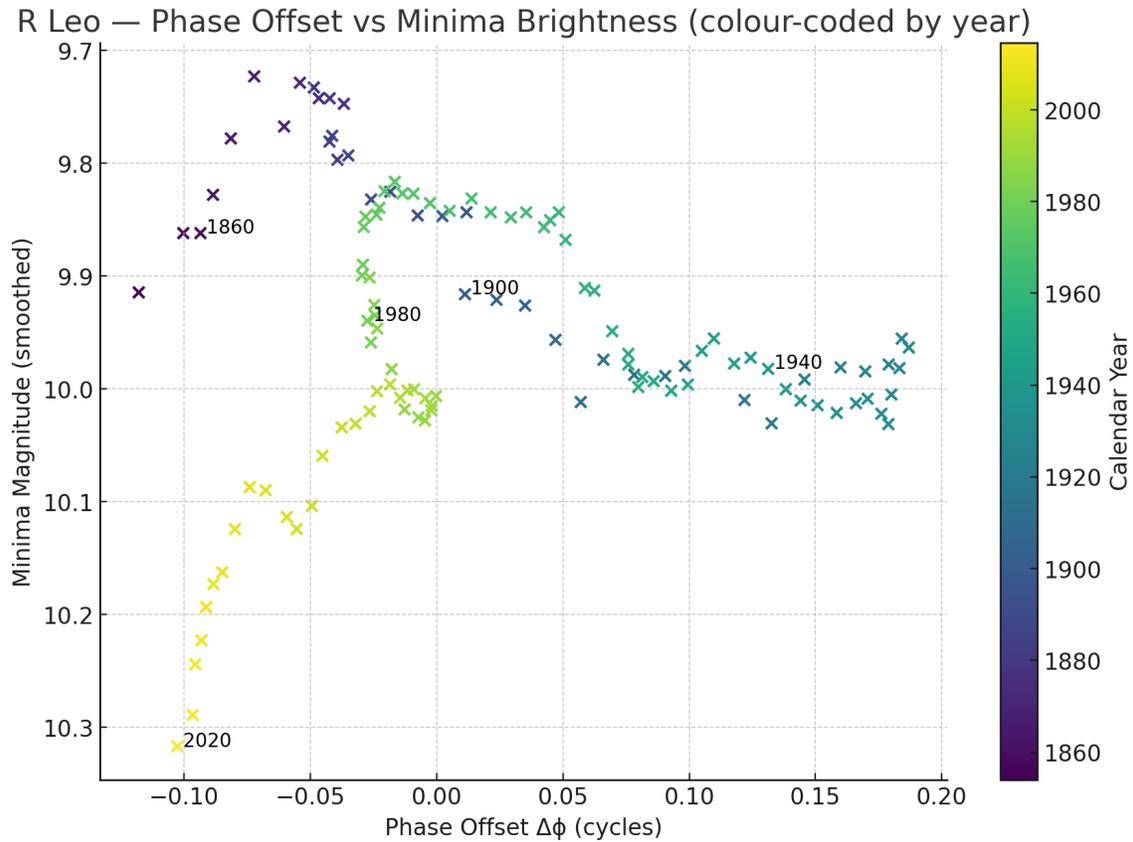

Figure 23. Phase offset, minimal brightness and date compared.

In neither case is there a sign of any consistent relationship between period change and magnitude change. Since changes in period are most likely related to modifications to the kappa mechanism by convective effects in the star's inner layers, the independence of the changes of magnitude suggests that different mechanisms are at play, perhaps related to circumstellar extinction.

# 6. CONCLUSIONS

Over the last two centuries, the mean period of R Leo has gradually shortened by about 3 days, a reduction of approximately 1%.

Superimposed on the mean ~312-day period are clear modulations on timescales of approximately 35 and 98 years (41 and 115 cycles). When a minimum occurs earlier or later than average, the following minimum also tends to be early or late, and this behaviour persists for up to about 12 years (14 cycles). A similar persistence is observed for maxima.

The distribution in time of the depths of minima is non-random. This is also true of maxima, but the statistical significance is lower.

For minima, although the pattern of depths of minima is not strictly periodic, there is a significant quasi-periodicity around 110 to 115 cycles (85 to 98 years).

Adjacent minima have a strong tendency to be similar in depth: the median difference is 0.20 magnitude, compared to a 0.31 median difference between random pairs. This effect is not found in maxima. This depth similarity between neighbouring minima persists for more distant neighbours; even minima separated by dozens of cycles still retain their depth similarity, on average. This coherence remains detectable for separations of up to about 50 cycles (43 years).

This correlation between neighbouring minima is an aspect of long term gradual changes in the mean depths of minima. These changes do not appear to be cyclic, nor to be due to changes in photospheric brightness, but may instead track variations in the optical depth of the circumstellar environment of R Leo.

Despite similarity in timescales, there is no obvious correlation between changes in period and in extrema brightnesses.

During the 1840s, mean minima brightness fell and rose again, reaching its highest recorded mean levels in the 1870s before a sharp diminution from about 1880 to about 1911. This was followed by a long period of relative stability until about 1980. Since then the average minima brightness has fallen.

Meanwhile, until about 1980, maxima brightnesses changed in approximately the same way, though with a smaller amplitude. Since 1980 this trend seems to have reversed and the steady fall in mean brightness at minimum has been matched by a modest gradual increase in the mean brightness of maxima.

A tentative account of these alterations in mean brightness in terms of changes in circumstellar dust is as follows:
*1. Adjacent minima are more similar in depth than random pairs.*

The underlying mechanism that governs the amount of dust produced at each minimum usually operates for more than one cycle length (see 3. below).

*2. Adjacent maxima are no more similar in depth than random pairs.*

Most of the dust formed near each minimum must be destroyed or dispersed before the following minimum. If a substantial fraction of the newly formed dust were to persist near its formation region, it would produce a rapid and progressive cycle-to-cycle decline in minimum light that is not observed. Even if dust contributes only 0.5 magnitude[3] of the ~4 magnitude decline from maximum to minimum, and if only half of that dust survived locally into the subsequent cycle, the resulting cumulative dimming would be 0.25 magnitude per cycle.

Such an effect would lead to a pronounced secular fading over only a few cycles. In fact, the amount of dust left over after each cycle, which is here assumed to lead to the gradual declines in light explained in 3. and 4. below, causes only about 0.1 reduction in mean maximum magnitude over 10 cycles even during the most rapid intervals of long-term change (figure 18).

*3. Gradual declines in the brightness of minima continue for up to about 50 cycles.*

Elevated dust production rates can be maintained for many cycles.

The persistence of elevated dust yield over many cycles implies a slow control mechanism acting on the dust-forming process, lasting for an order of magnitude longer than the ~312-day pulsation period. Candidate mechanisms discussed in the literature include intrinsic radiative–dust instabilities in dust-driven winds (e.g. the "exterior κ-mechanism"; Woitke 2006), slow variations in the extended-atmosphere background produced by convection–pulsation coupling in nonlinear wind models (Höfner & Olofsson 2018; Freytag & Höfner 2023), and longer-timescale variability phenomena such as long secondary periods that may modulate mass loss and dust obscuration in long-period variables (e.g. Pawlak 2022).

Because material formed near each minimum is not completely cleared during the subsequent maximum, the circumstellar optical depth increases on average over time. This cumulative increase progressively attenuates photospheric light, producing multi-cycle declines in minimum brightness. This is consistent with the work of Danchi et al (1994), who find that dust production is sporadic, with episodes of substantial dust production typically separated by a few decades.

---

[3] How much of the ~4 magnitude decline of R Leo each cycle is attributable specifically to dust is unknown, though it is evident that dust and gas together account for the majority of the dimming (see, for example Reid & Goldston 2002). Evidence that silicate dust can significantly attenuate visual light is provided by the oxygen-rich Mira-like variable L2 Pup, which underwent a ~2 magnitude reduction in mean visual magnitude that was attributed to dust formation along the line of sight, accompanied by strong 10-μm silicate emission (Bedding et al. 2002).

The amount of surviving dust required to account for the observed changes in mean minimum magnitude is relatively small:

A change of 0.1 magnitude at visual wavelengths corresponds to an optical-depth increment of $\Delta\tau \approx 0.09$, using the standard relation $A = 1.086\,\tau$.

If the excess extinction is entirely due to dust, then, adopting standard dust-grain parameters (grain radius $a \approx 0.1$ μm and extinction efficiency $Q \approx 1$), this implies an incremental dust column density of $\Delta\Sigma\_\text{dust} \approx 4 \times 10^{-6}$ g cm$^{-2}$.

Assuming spherical symmetry, and taking the characteristic radius of the near-photospheric dust reservoir to be $R = 2$–$3$ AU, the implied net dust mass increment is $\approx (2$–$5) \times 10^{-11}$ M$\odot$ per 0.1 magnitude of excess extinction.

A typical total mass-loss rate for a star such as R Leo is $\dot{M} \approx 1 \times 10^{-7}$ M$\odot$ yr$^{-1}$ (Fonfría et al. 2019). Assuming a gas-to-dust ratio of 200 gives a dust mass-loss rate of $\dot{M}_{\text{dust}} \approx 5 \times 10^{-10}$ M$\odot$ yr$^{-1}$.

Hence, if the ~0.3 decline in mean minimum magnitude between the 1880s and the early 1910s was indeed due to the net survival of dust, the total dust mass involved would be of order $10^{-10}$ M$\odot$, corresponding to only ~20% of the dust produced in a typical year. A modest imbalance between dust formation and removal is therefore sufficient to explain the observed secular dimming.

*4. The brightnesses of maxima change in the same direction as minima do, though with a smaller amplitude (a ~30% change in minimum-light flux is matched by only ~20% at maximum).*

Long-term variations that affect minimum brightness also influence maximum brightness, indicating a common underlying driver. The surviving circumstellar dust responsible for these changes is present at both minimum and maximum light. However, the same amount of surviving dust produces a larger fractional attenuation when the star is intrinsically faint than when it is intrinsically bright. At maximum light, when the photosphere is hotter and molecular opacity is reduced, identical amounts of dust therefore result in smaller fractional flux variations than at minimum light.

*5. R Leo's minima and maxima also undergo extended periods of gradual brightening, lasting up to ~75 cycles.*

When dust formation rates decline, they can remain suppressed for many cycles. During these intervals, excess circumstellar material produced during earlier high-loss states is gradually dissipated—through outward expansion, dilution, heating, or other processing—reducing its optical depth and allowing increasing amounts of photospheric light to escape.

*6. Since about 1980, the light curve has changed its character: each new minimum has tended to be deeper than the previous one while each new maximum has tended to be brighter than its predecessor.*

This finding cannot be explained within the same framework as the preceding behaviour. One possibility might be an increase in the proportion of molecular species to dust formed each cycle. Since molecular species are highly effective at blocking light at minimum, but also much more readily destroyed by post-minimum heating, that might lead both to deeper minima and brighter maxima. However, since there is neither a theoretical expectation of such an increase, nor any spectroscopic evidence of major change in the amounts of molecular species present, this is highly speculative.

*NOTE: To study the historical development of R Leo's circumstellar environment more directly, I am collating mid-infrared flux measurements from the mid-20th century to the present, with particular interest in N, Z, and Q band data. Information on relevant unpublished ground-based observations would be very welcome; please contact me at mikegoldsmith@gmail.com.*

# 7. APPENDICES

## Appendix A

Since R Leo is a distinctly red star at all phases, there is a risk that different observational photometric techniques – visual estimates, photography, photoelectric, CCD – may give different results even when the actual magnitude of the star is unchanged. This was checked by comparing the mean values of V magnitudes obtained by these methods. To ensure that the star was bright enough to be detected by each method, only observations brighter than magnitude 7 were considered. In case the actual mean magnitude of R Leo has changed since photometric records of the AAVSO began, data was binned into decades. The result (Figure 24) shows that different methods are in fact inconsistent.

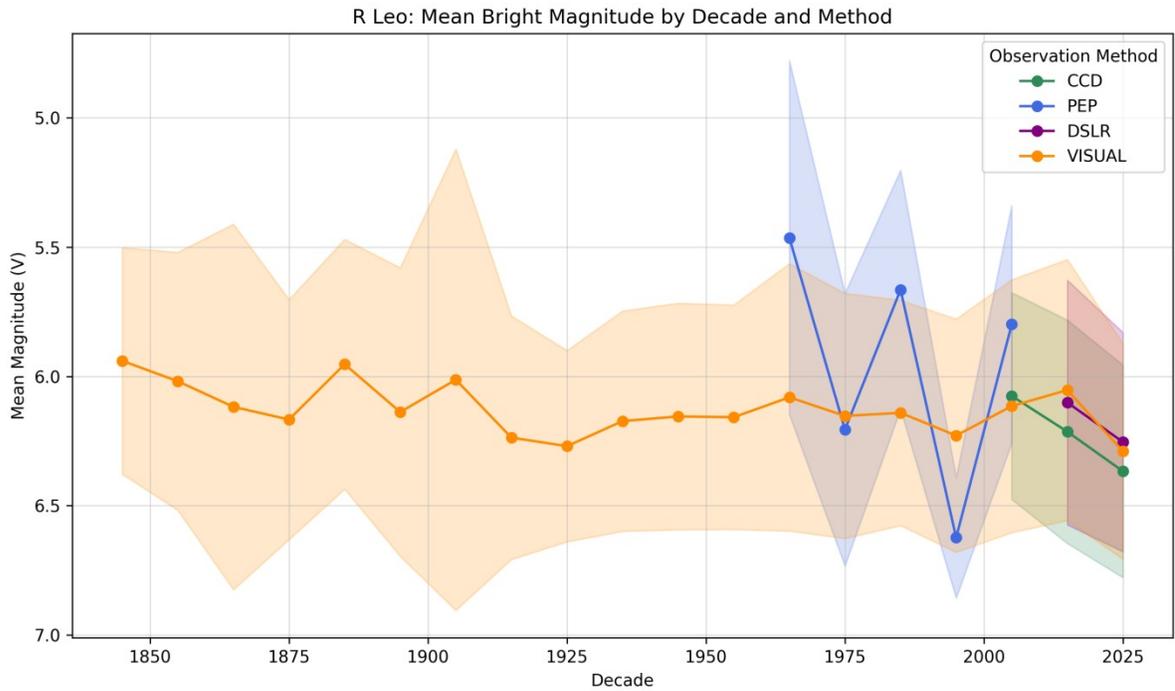

Figure 24 Comparison of mean V magnitudes obtained by different methods. Data points are means per decade, coloured bands are one sigma standard deviations.

Consequently the only observations used in this study are visual estimates.

# Appendix B

Historical papers with estimates of R Leo's maxima and minima are listed in Hoeppe 1982. In most cases these papers do not include individual observations, and most of those that do are represented in the AAVSO database. An exception is Pračka 1916.

This contains 434 observations of R Leo made by Prof. Adalbert (Vojtěch) Šafařík between 1877 and 1892, collated and checked by Ladislav Pračka. Plotting this data along with AAVSO values indicates that this additional data is not significantly biased (figure 25), and it was therefore included in the dataset analysed in the current paper.

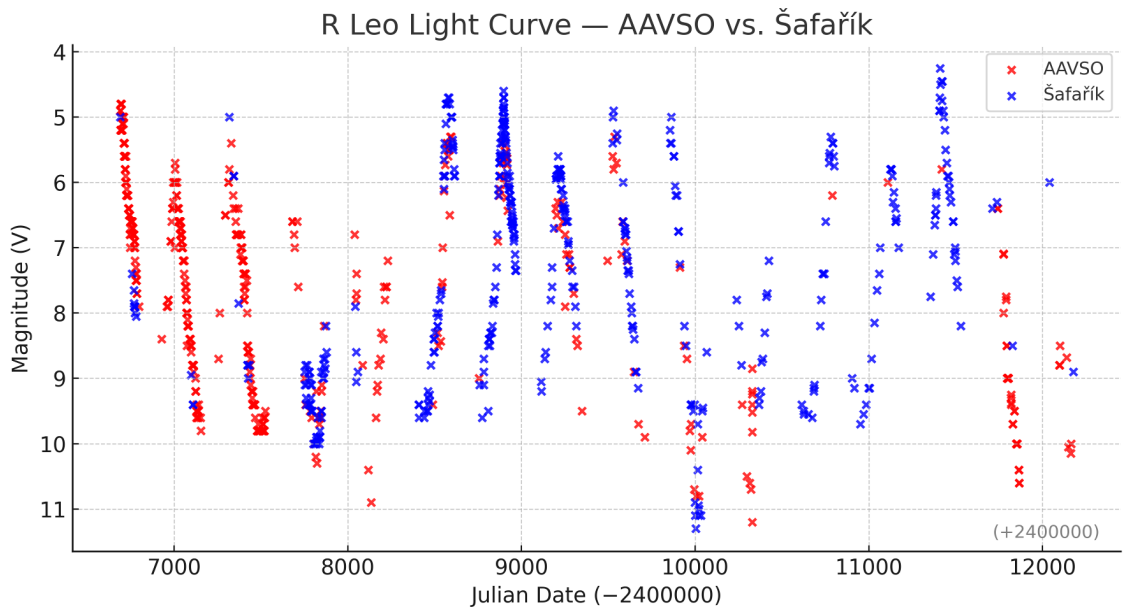

Figure 25. AAVSO data compared with Šafařík observations.

# Appendix C

Table 1. Maxima

| Cycle | Source | Date | Julian Date | Magnitude | Fit |
|---|---|---|---|---|---|
| 79 | Hoeppe | 1824-Dec-19 | 2387615 | 5.9 | Good |
| 83 | Hoeppe | 1828-May-19 | 2388862 | 6.4 | Good |
| 84 | Hoeppe | 1829-Mar-23 | 2389170 | 6.1 | Good |
| 85 | Hoeppe | 1830-Jan-26 | 2389479 | 6.0 | Good |
| 105 | AAVSO | 1847-Mar-16 | 2395737 | 5.1 | Good |
| 106 | AAVSO | 1848-Jan-17 | 2396044 | 5.5 | Good |
| 107 | AAVSO | 1848-Dec-21 | 2396383 | 5.3 | Good |
| 108 | Hoeppe | 1849-Oct-14 | 2396680 | 5.3 | Good |
| 111 | AAVSO | 1852-Apr-27 | 2397606 | 5.2 | Good |
| 112 | Hoeppe | 1853-Mar-13 | 2397926 | 5.2 | Good |
| 113 | AAVSO | 1854-Jan-26 | 2398245 | 5.3 | Rough |
| 114 | AAVSO | 1854-Dec-19 | 2398572 | 5.8 | Rough |
| 117 | AAVSO | 1857-Jun-27 | 2399493 | 5.3 | Good |
| 118 | AAVSO | 1858-Apr-29 | 2399799 | 6.1 | Good |
| 119 | AAVSO | 1859-Mar-07 | 2400111 | 5.3 | Good |
| 120 | AAVSO | 1860-Jan-14 | 2400424 | 6.1 | Good |
| 121 | AAVSO | 1860-Nov-27 | 2400742 | 5.8 | Good |
| 122 | AAVSO | 1861-Sep-24 | 2401043 | 6.0 | Rough |
| 124 | AAVSO | 1863-Jun-18 | 2401675 | 5.8 | Rough |
| 125 | Hoeppe | 1864-Apr-22 | 2401984 | 5.4 | Good |
| 126 | Hoeppe | 1865-Mar-04 | 2402300 | 6.3 | Good |
| 127 | Hoeppe | 1866-Jan-20 | 2402622 | 6.3 | Good |
| 128 | AAVSO | 1866-Dec-01 | 2402937 | 5.6 | Good |
| 129 | AAVSO | 1867-Sep-10 | 2403220 | 6.2 | Rough |
| 131 | Hoeppe | 1869-Jun-25 | 2403874 | 5.6 | Good |
| 132 | AAVSO | 1870-May-14 | 2404197 | 5.8 | Good |
| 133 | Hoeppe | 1871-Mar-16 | 2404503 | 5.9 | Good |
| 134 | AAVSO | 1872-Jan-22 | 2404815 | 5.6 | Good |
| 135 | Hoeppe | 1872-Dec-03 | 2405131 | 6.1 | Good |
| 136 | Hoeppe | 1873-Oct-18 | 2405450 | 6.4 | Good |
| 138 | Hoeppe | 1875-Jul-04 | 2406074 | 5.4 | Good |
| 139 | AAVSO | 1876-May-06 | 2406381 | 5.8 | Good |
| 140 | AAVSO & Šafařík | 1877-Mar-13 | 2406692 | 5.0 | Good |
| 141 | AAVSO & Šafařík | 1878-Jan-16 | 2407001 | 6.0 | Good |
| 142 | Hoeppe | 1878-Dec-04 | 2407323 | 5.8 | Good |
| 146 | Hoeppe | 1882-May-16 | 2408582 | 5.3 | Good |
| 147 | Hoeppe | 1883-Apr-02 | 2408903 | 4.8 | Good |
| 148 | AAVSO & Šafařík | 1884-Feb-12 | 2409219 | 5.8 | Rough |
| 149 | Hoeppe | 1884-Dec-20 | 2409531 | 5.4 | Good |
| 150 | AAVSO & Šafařík | 1885-Nov-10 | 2409856 | 5.4 | Rough |
| 153 | AAVSO & Šafařík | 1888-May-22 | 2410780 | 5.3 | Good |
| 154 | Hoeppe | 1889-Apr-15 | 2411108 | 5.7 | Good |
| 155 | AAVSO & Šafařík | 1890-Feb-13 | 2411412 | 4.5 | Good |
| 157 | Hoeppe | 1891-Nov-01 | 2412038 | 6.2 | Good |
| 159 | Hoeppe | 1893-Jul-06 | 2412651 | 5.7 | Good |
| 160 | Hoeppe | 1894-May-06 | 2412955 | 6.4 | Good |
| 161 | AAVSO | 1895-Mar-11 | 2413264 | 5.0 | Good |
| 162 | Hoeppe | 1896-Jan-09 | 2413568 | 6.4 | Good |
| 163 | Hoeppe | 1896-Nov-17 | 2413881 | 6.6 | Good |
| 166 | Hoeppe | 1899-Jun-07 | 2414813 | 6.2 | Good |
| 167 | AAVSO | 1900-Apr-06 | 2415116 | 5.3 | Good |
| 168 | AAVSO | 1901-Feb-19 | 2415435 | 6.0 | Good |
| 169 | Hoeppe | 1901-Dec-27 | 2415746 | 5.8 | Good |
| 170 | Hoeppe | 1902-Nov-06 | 2416060 | 6.6 | Good |
| 172 | Hoeppe | 1904-Jul-02 | 2416664 | 5.5 | Good |
| 173 | Hoeppe | 1905-May-17 | 2416983 | 5.9 | Good |
| 174 | AAVSO | 1906-Apr-05 | 2417306 | 5.5 | Good |
| 175 | AAVSO | 1907-Feb-07 | 2417614 | 5.3 | Good |
| 176 | AAVSO | 1907-Dec-15 | 2417925 | 5.9 | Good |
| 177 | Waagen | 1908-Nov-11 | 2418257 | 6.1 | Good |
| 178 | AAVSO | 1909-Sep-20 | 2418570 | 5.9 | Good |
| 179 | Hoeppe | 1910-Jul-22 | 2418875 | 5.9 | Good |
| 180 | Hoeppe | 1911-Jun-25 | 2419213 | 6.0 | Good |
| 181 | AAVSO | 1912-Apr-28 | 2419521 | 5.2 | Good |
| 182 | AAVSO | 1913-Mar-18 | 2419845 | 6.5 | Good |
| 183 | AAVSO | 1914-Feb-12 | 2420176 | 6.0 | Good |
| 184 | Hoeppe | 1914-Dec-24 | 2420491 | 6.1 | Good |
| 185 | Hoeppe | 1915-Nov-11 | 2420813 | 5.5 | Good |
| 186 | AAVSO | 1916-Sep-13 | 2421120 | 5.4 | Good |
| 187 | Hoeppe | 1917-Jul-11 | 2421421 | 6.2 | Good |
| 188 | Hoeppe | 1918-Jun-05 | 2421750 | 5.7 | Good |
| 189 | Waagen | 1919-Apr-15 | 2422064 | 6.9 | Good |
| 190 | AAVSO | 1920-Mar-29 | 2422413 | 6.1 | Good |
| 191 | AAVSO | 1920-Dec-31 | 2422690 | 5.8 | Good |
| 192 | AAVSO | 1921-Nov-30 | 2423024 | 5.8 | Good |
| 193 | AAVSO | 1922-Oct-14 | 2423342 | 5.9 | Good |
| 195 | AAVSO | 1924-Jun-12 | 2423949 | 5.8 | Good |
| 196 | AAVSO | 1925-Apr-20 | 2424261 | 6.0 | Good |
| 197 | AAVSO | 1926-Feb-19 | 2424566 | 5.5 | Good |
| 198 | AAVSO | 1927-Jan-05 | 2424886 | 5.8 | Good |
| 199 | AAVSO | 1927-Nov-12 | 2425197 | 5.8 | Good |
| 200 | AAVSO | 1928-Sep-06 | 2425496 | 5.8 | Good |
| 201 | Hoeppe | 1929-Aug-14 | 2425838 | 5.9 | Good |
| 202 | AAVSO | 1930-Jun-22 | 2426150 | 5.6 | Good |
| 203 | AAVSO | 1931-Apr-25 | 2426457 | 5.5 | Good |
| 204 | AAVSO | 1932-Mar-13 | 2426780 | 5.9 | Good |
| 205 | AAVSO | 1933-Jan-15 | 2427088 | 5.9 | Good |
| 206 | AAVSO | 1933-Oct-24 | 2427370 | 5.8 | Good |
| 207 | Hoeppe | 1934-Sep-03 | 2427684 | 5.7 | Good |
| 208 | Hoeppe | 1935-Jul-08 | 2427992 | 5.3 | Good |
| 209 | Hoeppe | 1936-May-20 | 2428309 | 5.6 | Good |
| 210 | AAVSO | 1937-Mar-23 | 2428616 | 5.4 | Good |
| 211 | AAVSO | 1938-Jan-24 | 2428923 | 5.6 | Good |
| 212 | Waagen | 1938-Nov-18 | 2429221 | 5.5 | Good |
| 213 | AAVSO | 1939-Oct-04 | 2429541 | 5.6 | Good |
| 215 | AAVSO | 1941-Jun-08 | 2430154 | 5.4 | Good |
| 216 | AAVSO | 1942-Apr-27 | 2430477 | 6.5 | Good |
| 217 | AAVSO | 1943-Mar-11 | 2430795 | 6.2 | Good |
| 218 | AAVSO | 1944-Jan-09 | 2431099 | 6.0 | Good |
| 219 | Waagen | 1944-Nov-05 | 2431400 | 5.7 | Fair |
| 222 | AAVSO | 1947-May-31 | 2432337 | 5.4 | Good |
| 223 | AAVSO | 1948-Mar-26 | 2432637 | 5.7 | Good |
| 224 | AAVSO | 1949-Feb-07 | 2432955 | 6.0 | Good |
| 225 | AAVSO | 1949-Dec-22 | 2433273 | 5.3 | Good |
| 227 | AAVSO | 1951-Sep-20 | 2433910 | 5.4 | Rough |
| 229 | AAVSO | 1953-May-29 | 2434527 | 5.2 | Good |
| 230 | AAVSO | 1954-Apr-11 | 2434844 | 6.0 | Good |
| 231 | AAVSO | 1955-Feb-12 | 2435151 | 5.7 | Good |
| 232 | AAVSO | 1956-Jan-02 | 2435475 | 5.9 | Good |
| 233 | AAVSO | 1956-Nov-09 | 2435787 | 6.1 | Good |
| 234 | Waagen | 1957-Sep-12 | 2436094 | 5.6 | Rough |
| 235 | Waagen | 1958-Jul-06 | 2436391 | 5.0 | Rough |
| 236 | AAVSO | 1959-May-23 | 2436712 | 5.5 | Good |
| 237 | AAVSO | 1960-Mar-28 | 2437022 | 5.2 | Good |
| 238 | AAVSO | 1961-Feb-05 | 2437336 | 5.4 | Good |
| 239 | AAVSO | 1961-Dec-24 | 2437658 | 5.5 | Good |
| 240 | Waagen | 1962-Oct-22 | 2437960 | 5.9 | Good |
| 241 | Waagen | 1963-Aug-28 | 2438270 | 5.8 | Rough |
| 242 | Waagen | 1964-Jul-06 | 2438583 | 5.5 | Fair |
| 243 | AAVSO | 1965-May-14 | 2438895 | 5.2 | Good |
| 244 | AAVSO | 1966-Mar-16 | 2439201 | 6.0 | Good |
| 245 | AAVSO | 1967-Jan-16 | 2439507 | 5.5 | Good |
| 246 | AAVSO | 1967-Nov-30 | 2439825 | 5.8 | Good |
| 247 | Waagen | 1968-Sep-24 | 2440124 | 5.9 | Good |
| 248 | Waagen | 1969-Aug-04 | 2440438 | 5.8 | Good |
| 249 | AAVSO | 1970-May-30 | 2440737 | 5.4 | Good |
| 250 | AAVSO | 1971-Apr-19 | 2441061 | 6.3 | Good |
| 251 | Waagen | 1972-Feb-21 | 2441369 | 5.9 | Good |
| 252 | AAVSO | 1973-Jan-02 | 2441685 | 6.4 | Good |
| 253 | AAVSO | 1973-Nov-04 | 2441991 | 5.6 | Good |
| 254 | Waagen | 1974-Sep-12 | 2442303 | 6.0 | Good |
| 255 | Waagen | 1975-Jul-16 | 2442610 | 6.0 | Fair |
| 256 | AAVSO | 1976-May-16 | 2442915 | 6.0 | Good |
| 257 | AAVSO | 1977-Mar-18 | 2443221 | 5.4 | Good |
| 258 | AAVSO | 1978-Feb-05 | 2443545 | 6.3 | Good |
| 259 | AAVSO | 1978-Dec-02 | 2443845 | 5.2 | Good |
| 260 | AAVSO | 1979-Oct-10 | 2444157 | 5.5 | Good |
| 261 | Waagen | 1980-Aug-17 | 2444469 | 5.9 | Fair |
| 262 | Waagen | 1981-Jul-02 | 2444788 | 5.5 | Good |
| 263 | AAVSO | 1982-May-21 | 2445111 | 5.3 | Good |
| 264 | AAVSO | 1983-Apr-10 | 2445435 | 5.8 | Good |
| 265 | AAVSO | 1984-Feb-03 | 2445734 | 6.1 | Good |
| 266 | AAVSO | 1984-Dec-30 | 2446065 | 5.8 | Good |
| 267 | AAVSO | 1985-Nov-13 | 2446383 | 5.4 | Good |
| 268 | Waagen | 1986-Sep-12 | 2446686 | 6.2 | Good |
| 269 | Waagen | 1987-Jul-15 | 2446992 | 5.8 | Good |
| 270 | AAVSO | 1988-May-25 | 2447307 | 5.9 | Good |
| 271 | AAVSO | 1989-Apr-13 | 2447630 | 5.5 | Good |
| 272 | AAVSO | 1990-Feb-20 | 2447943 | 6.1 | Good |
| 273 | AAVSO | 1990-Dec-23 | 2448249 | 5.9 | Good |
| 274 | AAVSO | 1991-Oct-26 | 2448556 | 5.1 | Good |
| 277 | AAVSO | 1994-May-23 | 2449496 | 5.3 | Good |
| 278 | AAVSO | 1995-Apr-01 | 2449809 | 6.5 | Good |
| 279 | AAVSO | 1996-Feb-07 | 2450121 | 5.7 | Good |
| 280 | AAVSO | 1996-Dec-21 | 2450439 | 6.4 | Good |
| 281 | AAVSO | 1997-Oct-22 | 2450744 | 6.1 | Good |
| 283 | AAVSO | 1999-Jun-27 | 2451357 | 5.5 | Good |
| 284 | AAVSO | 2000-May-04 | 2451669 | 5.9 | Good |
| 285 | AAVSO | 2001-Mar-06 | 2451975 | 5.3 | Good |
| 286 | AAVSO | 2002-Jan-12 | 2452287 | 5.7 | Good |
| 287 | AAVSO | 2002-Nov-26 | 2452605 | 5.7 | Good |
| 288 | AAVSO | 2003-Sep-28 | 2452911 | 5.9 | Good |
| 290 | AAVSO | 2005-Jul-08 | 2453560 | 5.2 | Rough |
| 291 | AAVSO | 2006-Apr-21 | 2453847 | 5.5 | Good |
| 292 | AAVSO | 2007-Feb-26 | 2454158 | 5.9 | Good |
| 293 | AAVSO | 2008-Jan-05 | 2454471 | 5.4 | Good |
| 294 | AAVSO | 2008-Nov-11 | 2454782 | 5.3 | Good |
| 297 | AAVSO | 2011-May-19 | 2455701 | 5.2 | Good |
| 298 | AAVSO | 2012-Apr-07 | 2456025 | 6.1 | Good |

| 299 | AAVSO | 2013-Feb-07 | 2456331 | 5.8 | Good |
| 300 | AAVSO | 2013-Dec-16 | 2456643 | 5.4 | Good |
| 301 | AAVSO | 2014-Nov-09 | 2456971 | 5.4 | Rough |
| 304 | AAVSO | 2017-May-23 | 2457897 | 6.1 | Good |
| 305 | AAVSO | 2018-Mar-24 | 2458202 | 5.2 | Good |
| 306 | AAVSO | 2019-Feb-12 | 2458527 | 5.5 | Good |
| 307 | AAVSO | 2019-Dec-21 | 2458839 | 5.5 | Good |
| 308 | AAVSO | 2020-Nov-08 | 2459162 | 5.7 | Good |
| 310 | AAVSO | 2022-Jul-08 | 2459769 | 5.0 | Rough |
| 311 | AAVSO | 2023-May-10 | 2460075 | 6.4 | Good |
| 312 | AAVSO | 2024-Mar-17 | 2460387 | 6.0 | Good |
| 313 | AAVSO | 2025-Jan-29 | 2460705 | 5.7 | Good |

Table 2. Minima

| Cycle | Source | Date | Julian Date | Magnitude | Fit |
|---|---|---|---|---|---|
| 79 | Hoeppe | 1825-Jun-04 | 2387782 | 10.0 | Good |
| 80 | Hoeppe | 1826-Mar-31 | 2388082 | 9.8 | Good |
| 82 | Hoeppe | 1828-Jan-03 | 2388725 | 10.5 | Good |
| 83 | Hoeppe | 1828-Oct-29 | 2389025 | 10.6 | Good |
| 86 | Hoeppe | 1831-May-19 | 2389957 | 10.1 | Good |
| 87 | Hoeppe | 1832-Mar-16 | 2390259 | 10.0 | Good |
| 108 | AAVSO | 1850-Mar-22 | 2396839 | 10.8 | Good |
| 115 | AAVSO | 1856-Mar-05 | 2399014 | 9.9 | Rough |
| 116 | AAVSO | 1857-Jan-26 | 2399341 | 10.1 | Good |
| 119 | Hoeppe | 1859-Sep-18 | 2400306 | 9.5 | Good |
| 122 | AAVSO | 1862-Mar-24 | 2401224 | 9.5 | Good |
| 124 | AAVSO | 1863-Dec-01 | 2401841 | 9.4 | Good |
| 125 | Hoeppe | 1864-Nov-04 | 2402180 | 10.1 | Good |
| 126 | Hoeppe | 1865-Sep-09 | 2402489 | 9.8 | Good |
| 128 | AAVSO | 1867-May-02 | 2403089 | 9.8 | Good |
| 129 | AAVSO | 1868-Mar-07 | 2403399 | 9.7 | Good |
| 130 | Hoeppe | 1869-Jan-20 | 2403718 | 9.6 | Good |
| 135 | AAVSO | 1873-May-23 | 2405302 | 9.6 | Good |
| 136 | AAVSO | 1874-Mar-30 | 2405613 | 9.5 | Good |
| 137 | AAVSO | 1875-Feb-13 | 2405933 | 9.9 | Good |
| 138 | AAVSO | 1876-Mar-07 | 2406321 | 8.9 | Rough |
| 141 | Hoeppe | 1878-Jul-12 | 2407178 | 9.8 | Good |
| 142 | AAVSO & Šafařík | 1879-May-27 | 2407497 | 9.8 | Good |
| 145 | Hoeppe | 1881-Dec-20 | 2408435 | 9.6 | Good |
| 146 | Hoeppe | 1882-Nov-06 | 2408756 | 9.0 | Good |
| 151 | AAVSO & Šafařík | 1887-Feb-21 | 2410324 | 10.9 | Rough |
| 158 | Hoeppe | 1893-Feb-01 | 2412496 | 10.0 | Good |
| 160 | Hoeppe | 1894-Oct-15 | 2413117 | 10.2 | Good |
| 161 | Hoeppe | 1895-Sep-22 | 2413459 | 10.2 | Good |
| 162 | AAVSO | 1896-Apr-13 | 2413663 | 9.2 | Rough |
| 164 | AAVSO | 1898-Apr-02 | 2414382 | 9.7 | Good |
| 170 | AAVSO | 1903-Apr-24 | 2416229 | 10.5 | Good |
| 171 | AAVSO | 1904-Feb-18 | 2416529 | 9.8 | Good |
| 177 | AAVSO | 1909-Apr-28 | 2418425 | 9.7 | Good |
| 178 | AAVSO | 1910-Mar-16 | 2418747 | 10.1 | Good |
| 179 | AAVSO | 1911-Feb-05 | 2419073 | 10.5 | Good |
| 180 | AAVSO | 1911-Dec-12 | 2419383 | 9.5 | Good |
| 184 | AAVSO | 1915-Jun-26 | 2420675 | 10.0 | Rough |
| 185 | AAVSO | 1916-Apr-25 | 2420979 | 9.5 | Good |
| 186 | AAVSO | 1917-Mar-10 | 2421298 | 9.9 | Good |
| 188 | AAVSO | 1918-Dec-07 | 2421935 | 10.4 | Good |
| 192 | AAVSO | 1922-May-11 | 2423186 | 9.9 | Good |
| 193 | AAVSO | 1923-Mar-19 | 2423498 | 9.9 | Good |
| 194 | AAVSO | 1924-Jan-31 | 2423816 | 10.2 | Good |
| 196 | AAVSO | 1925-Oct-12 | 2424436 | 10.1 | Good |
| 198 | AAVSO | 1927-Jun-28 | 2425060 | 10.1 | Good |
| 199 | AAVSO | 1928-May-11 | 2425378 | 10.3 | Good |
| 200 | AAVSO | 1929-Mar-17 | 2425688 | 10.2 | Good |
| 201 | AAVSO | 1930-Jan-25 | 2426002 | 10.0 | Good |
| 202 | AAVSO | 1930-Nov-23 | 2426304 | 9.8 | Good |
| 203 | AAVSO | 1931-Oct-16 | 2426631 | 10.1 | Good |
| 205 | AAVSO | 1933-Jun-19 | 2427243 | 9.7 | Good |
| 206 | AAVSO | 1934-Apr-22 | 2427550 | 9.6 | Good |
| 207 | AAVSO | 1935-Feb-23 | 2427857 | 9.8 | Good |
| 208 | AAVSO | 1936-Jan-07 | 2428175 | 10.0 | Good |
| 209 | AAVSO | 1936-Nov-03 | 2428476 | 10.0 | Good |
| 212 | AAVSO | 1939-May-12 | 2429396 | 9.6 | Good |
| 213 | AAVSO | 1940-Mar-21 | 2429710 | 10.4 | Good |
| 214 | AAVSO | 1941-Jan-22 | 2430017 | 10.0 | Good |
| 215 | AAVSO | 1941-Dec-20 | 2430349 | 10.4 | Good |
| 216 | AAVSO | 1942-Oct-30 | 2430663 | 10.2 | Good |
| 218 | AAVSO | 1944-Jul-01 | 2431273 | 9.6 | Good |
| 219 | AAVSO | 1945-May-14 | 2431590 | 10.0 | Good |
| 220 | AAVSO | 1946-Mar-09 | 2431889 | 10.4 | Good |
| 221 | AAVSO | 1947-Jan-12 | 2432198 | 10.0 | Good |
| 222 | AAVSO | 1947-Nov-25 | 2432515 | 10.1 | Good |
| 223 | AAVSO | 1948-Oct-13 | 2432838 | 10.1 | Good |
| 226 | AAVSO | 1951-Apr-28 | 2433765 | 9.9 | Good |
| 227 | AAVSO | 1952-Mar-08 | 2434080 | 9.8 | Good |
| 228 | AAVSO | 1953-Jan-07 | 2434385 | 9.9 | Good |
| 233 | AAVSO | 1957-Apr-22 | 2435951 | 9.7 | Good |
| 234 | AAVSO | 1958-Mar-08 | 2436271 | 9.9 | Good |
| 235 | AAVSO | 1959-Jan-05 | 2436574 | 10.2 | Good |
| 236 | AAVSO | 1959-Nov-01 | 2436874 | 9.9 | Good |
| 239 | AAVSO | 1962-May-29 | 2437814 | 9.8 | Good |
| 240 | AAVSO | 1963-Apr-10 | 2438130 | 10.0 | Good |
| 241 | AAVSO | 1964-Feb-18 | 2438444 | 9.8 | Good |
| 242 | AAVSO | 1964-Dec-13 | 2438743 | 10.0 | Good |
| 243 | AAVSO | 1965-Nov-11 | 2439076 | 9.8 | Good |
| 245 | Waagen | 1967-Jul-08 | 2439680 | 10.0 | Rough |
| 246 | AAVSO | 1968-May-13 | 2439990 | 9.5 | Good |
| 247 | AAVSO | 1969-Mar-27 | 2440308 | 9.6 | Good |
| 248 | AAVSO | 1970-Jan-13 | 2440600 | 9.1 | Good |
| 249 | AAVSO | 1970-Dec-09 | 2440930 | 9.9 | Good |
| 250 | AAVSO | 1971-Oct-09 | 2441234 | 10.1 | Good |
| 252 | AAVSO | 1973-Jul-06 | 2441870 | 10.2 | Good |
| 253 | AAVSO | 1974-Apr-27 | 2442165 | 9.8 | Good |
| 254 | AAVSO | 1975-Mar-01 | 2442473 | 10.0 | Good |
| 255 | AAVSO | 1975-Dec-29 | 2442776 | 9.7 | Good |
| 256 | AAVSO | 1976-Oct-31 | 2443083 | 9.7 | Good |
| 258 | Waagen | 1978-Jul-16 | 2443706 | 9.9 | Fair |
| 259 | AAVSO | 1979-Jun-06 | 2444031 | 9.8 | Good |
| 260 | AAVSO | 1980-Apr-12 | 2444342 | 10.0 | Good |
| 261 | AAVSO | 1981-Feb-20 | 2444656 | 9.9 | Good |
| 262 | AAVSO | 1982-Jan-10 | 2444980 | 9.6 | Good |
| 263 | AAVSO | 1982-Nov-15 | 2445289 | 10.1 | Good |
| 264 | AAVSO | 1983-Oct-08 | 2445616 | 10.1 | Good |
| 265 | Waagen | 1984-Aug-07 | 2445920 | 10.1 | Fair |
| 266 | AAVSO | 1985-Jun-16 | 2446233 | 9.9 | Good |
| 267 | AAVSO | 1986-Apr-26 | 2446547 | 10.2 | Good |
| 268 | AAVSO | 1987-Mar-04 | 2446859 | 10.2 | Good |
| 269 | AAVSO | 1988-Jan-16 | 2447177 | 9.8 | Good |
| 270 | AAVSO | 1988-Nov-19 | 2447485 | 10.0 | Good |
| 271 | AAVSO | 1989-Oct-11 | 2447811 | 10.3 | Rough |
| 273 | AAVSO | 1991-Jun-21 | 2448429 | 9.6 | Good |
| 274 | AAVSO | 1992-Apr-15 | 2448728 | 9.5 | Good |
| 275 | AAVSO | 1993-Mar-01 | 2449048 | 10.3 | Good |
| 276 | AAVSO | 1993-Dec-30 | 2449352 | 10.4 | Good |
| 277 | AAVSO | 1994-Nov-26 | 2449683 | 10.5 | Good |
| 278 | AAVSO | 1995-Sep-26 | 2449987 | 10.4 | Good |
| 280 | AAVSO | 1997-Jun-12 | 2450612 | 10.0 | Good |
| 281 | AAVSO | 1998-Apr-15 | 2450919 | 9.8 | Good |
| 282 | AAVSO | 1999-Feb-20 | 2451230 | 9.7 | Good |
| 283 | AAVSO | 1999-Dec-25 | 2451538 | 9.7 | Good |
| 284 | AAVSO | 2000-Oct-22 | 2451840 | 9.9 | Good |
| 286 | AAVSO | 2002-Jul-02 | 2452458 | 9.9 | Rough |
| 287 | AAVSO | 2003-May-12 | 2452772 | 9.9 | Good |
| 288 | AAVSO | 2004-Mar-23 | 2453088 | 10.1 | Good |
| 289 | AAVSO | 2005-Jan-28 | 2453399 | 10.0 | Good |
| 290 | AAVSO | 2005-Dec-01 | 2453706 | 9.9 | Good |
| 291 | AAVSO | 2006-Oct-02 | 2454011 | 10.3 | Rough |
| 293 | AAVSO | 2008-Jul-08 | 2454656 | 10.1 | Good |
| 294 | AAVSO | 2009-Apr-25 | 2454947 | 10.2 | Good |
| 295 | AAVSO | 2010-Mar-17 | 2455273 | 10.3 | Good |
| 296 | AAVSO | 2011-Jan-03 | 2455565 | 10.1 | Good |
| 297 | AAVSO | 2011-Nov-20 | 2455886 | 10.4 | Good |
| 298 | AAVSO | 2012-Sep-24 | 2456195 | 10.7 | Rough |

| 300 | AAVSO | 2014-Jun-15 | 2456824 | 10.2 | Good |
| 301 | AAVSO | 2015-Apr-14 | 2457127 | 10.0 | Good |
| 302 | AAVSO | 2016-Feb-25 | 2457444 | 10.3 | Good |
| 303 | AAVSO | 2017-Jan-05 | 2457759 | 10.8 | Good |
| 304 | AAVSO | 2017-Nov-14 | 2458072 | 10.6 | Good |
| 305 | AAVSO | 2018-Sep-30 | 2458392 | 9.9 | Rough |
| 307 | AAVSO | 2020-Jun-16 | 2459017 | 10.1 | Good |
| 308 | AAVSO | 2021-Apr-20 | 2459325 | 10.5 | Good |
| 309 | AAVSO | 2022-Feb-16 | 2459627 | 10.4 | Good |
| 310 | AAVSO | 2023-Jan-04 | 2459949 | 10.8 | Good |
| 311 | AAVSO | 2023-Oct-16 | 2460234 | 10.7 | Rough |